\documentclass{jfm}
\usepackage{amsmath,amsfonts,amssymb,amsgen,amsbsy,upmath,bm,graphicx,natbib}
\newcommand{\ud}{\mathrm{d}}
\newcommand{\ue}{\mathrm{e}}
\newcommand{\ui}{\mathrm{i}}
\newcommand{\mea}{\int_0^{\mathcal{T}}\!\frac{\ud t}{\mathcal{T}}\int_{\mathbb{B}}\!\ud\bm{x}\int_{\mathbb{R}^{d}}\!\ud\bm{v}}
\newcommand{\mear}{\int_0^{\mathcal{T}}\!\frac{\ud t}{\mathcal{T}}\int_{\mathbb{B}}\!\ud\bm{x}}

\title[Eddy~diffusivities of~inertial~particles under~gravity]
{Eddy~diffusivities of~inertial~particles under~gravity}

\author[M.\ Martins Afonso, A.\ Mazzino and P.\ Muratore-Ginanneschi]
 {M\ls A\ls R\ls C\ls O\ns M\ls A\ls R\ls T\ls I\ls N\ls S\ns A\ls F\ls O\ls N\ls S\ls O$^1$,
 A\ls N\ls D\ls R\ls E\ls A\ns M\ls A\ls Z\ls Z\ls I\ls N\ls O$^2$\ns\linebreak
 \and P\ls A\ls O\ls L\ls O\ns  M\ls U\ls R\ls A\ls T\ls O\ls R\ls E\ls-\ls G\ls I\ls N\ls A\ls N\ls N\ls E\ls S\ls C\ls H\ls I$^3$}
\affiliation{$^1$Universit\'e de Toulouse, INP/UPS/CNRS, Institut~de~M\'ecanique~des~Fluides~de~Toulouse~-~groupe~Particules~Spray~et~Combustion,
 all\'ee~du~Professeur~Camille~Soula, 31400~Toulouse, France
\\[\affilskip]$^2$Department~of~Physics~-~University~of~Genova, and CNISM \& INFN~-~Genova~Section, via~Dodecaneso~33, 16146~Genova, Italy
\\[\affilskip]$^3$Department~of~Mathematics~and~Statistics~–~University~of~Helsinki, PO~Box~4, 00014~Helsinki, Finland}
\date{\today}

\begin{document}
\maketitle

 \begin{abstract}
  The large-scale/long-time transport of inertial particles of arbitrary
  mass density under gravity is investigated by means of a formal multiple-scale perturbative expansion
  in the scale-separation parametre between the carrier flow and the particle
  concentration field. The resulting large-scale equation for the particle
  concentration is determined, and is found to be
  diffusive with a positive-definite eddy diffusivity. The calculation of the
  latter tensor is reduced to the resolution of an auxiliary differential problem, consisting
  of a coupled set of two differential equations in a $(6+1)$-dimensional
  coordinate system ($3$ space coordinates plus $3$ velocity coordinates plus time).
  Although expensive, numerical methods can be exploited to obtain the
  eddy diffusivity, for any desirable non-perturbative limit (e.g.\ arbitrary
  Stokes and Froude numbers). The aforementioned large-scale
  equation is then specialized to deal with two different relevant perturbative limits: i)
  vanishing of both Stokes time and sedimenting particle velocity; ii) vanishing
  Stokes time and finite sedimenting particle velocity. Both asymptotics
  lead to a greatly simplified auxiliary differential problem, now involving only
  space coordinates and thus easy to be tackled by standard numerical techniques.
  Explicit, exact expressions for the eddy diffusivities have been calculated,
  for both asymptotics, for the class of parallel flows, both static and
  time-dependent. This allows us to investigate analytically the role of
  gravity and inertia on the diffusion process by varying relevant features of the carrier
  flow, as e.g.\ the form of its temporal correlation function.
  Our results exclude a universal role played by gravity and inertia on the diffusive
  behaviour: regimes of both enhanced and reduced diffusion may exist,
  depending on the detailed structure of the carrier flow.
 \end{abstract}

\section{Introduction}

Transport of passive particles is a problem of major importance
in a variety of domains ranging from astrophysics and geophysics to
technological applications and biology \citep{RMP04}. A quantity of particular interest is the
rate at which particles are transported by the flow.

For inertialess particles (i.e.\ fluid particles),
for times large compared to those characteristic of the given velocity
field, transport is usually diffusive and characterized by effective (enhanced)
diffusion coefficients \citep{K87}, the so-called \emph{eddy diffusivities}.
These coefficients incorporate all the dynamical effects played by the velocity field on the particle transport.\\
Although the diffusive scenario with eddy diffusivities is
the typical one, there exist cases where superdiffusion is observed already for simple
incompressible laminar flows \citep{CMMV99} and synthetic flows
\citep{ACMV00}.\\
From an applicative point of view, eddy diffusivities have
long been fruitful concepts in turbulent-transport theory,
and their use has made the computation of turbulent-transport problems possible
at P\'eclet numbers too high for full numerical simulations
\citep[see, e.g.,][]{L97}.\\
In an asymptotic case of interaction between modes whose space and time
scales are strongly separated, eddy diffusivities have been calculated
exploiting a multiple-scale expansion in the scale parametre
by \citet{BCVV95}. Conversely,
for realistic flows, active on all space-time scales,
eddy diffusivities are generally dependent on all flow
characteristics and no general expression for them is known \citep{K87}.

The situation is even more complex in the case of transport of particles
with inertia \citep{G83,MR83,BFF01,WM03,CBBBCLMT06,VCVLMPT08}.
The main difficulty arises from the fact that in this case the fluid velocity
does not coincide with the particle velocity, with the result that a larger
phase space has to be considered (i.e.\ involving both particle position and
velocity). The standard Fokker--Planck equation
(which only involves space variables),
holding for the concentration field in the
case of fluid particles, is replaced, for particle with inertia,
by a Kramers equation for the phase-space density now involving
both space and velocity coordinates.\\
In many cases of interest, the dependence of the phase-space particle
density on the velocity coordinates is not of particular interest
(a relevant example is in the realm of pollutants dispersion,
where the crucial objective is to predict the space-time behaviour
of the pollutant concentration rather than the velocity distribution of the pollutant)
and can be averaged out in a way such that only space-time variables are involved.\\
Closed equations for such an observable (the so-called
marginal density) have been obtained by \citet{PS05} exploiting
a formal multiple-scale expansion in the scale-separation parametre. The attention
was focused there on the sole case of heavy particles and in the absence
of gravity. The latter is however relevant in many situations
of interest, including those where the particle acceleration is (at least)
of the same order of magnitude of the gravitational acceleration
\citep[see, e.g.,][]{M87}. This latter case corresponds to finite
terminal sedimenting velocity even in the presence of particles with
small inertia. A question which naturally arises is on whether
a constant (gravity-induced) falling/rising terminal velocity is
accompanied by a correction to the diffusion process occurring in the absence
of gravity. Answering this question is one of the main contribution
of the present paper.

To achieve our aim we have exploited a formal multiple-scale expansion
in the scale-separation parametre on the Kramers equation
for the particle probability density function.
Different particle densities have been considered,
ranging from the limit of heavy particles to the one of light
particles (i.e.\ bubbles). The spirit of the work is similar
to the one which motivated \citet{BCVV95}. Unfortunately,
for the reasons already mentioned, the analytical treatment is here
much more cumbersome, a fact which imposes a more formal treatment
of the multiple-scale expansion. With the hope of making the present
manuscript more readable, we decided to postpone
the most technical issues in dedicated appendices.
Here is a short summary of our main achievements.\\
We found that the large-scale/long-time behaviour of the particle
concentration field is diffusive and the eddy diffusivity is positive defined.
This holds for all Stokes times and particles density.
To obtain the eddy-diffusivity tensor one has to solve (numerically)
a system of two, one-way coupled, partial differential equations (the auxiliary problem) in a phase space involving both the particle coordinates
and their velocity (plus time). In a three-dimensional space, the differential problem
thus involves $7$ coordinates. Consequently, the numerical solution requires
a considerable computational effort.\\
In order to simplify the problem, and at the same time motivated by
situations of interests where the Stokes time is small, a further
perturbative analysis has been carried out in the limit of small inertia. Two asymptotics have been considered: one corresponding to
small Stokes and vanishing terminal sedimenting velocity, the other
corresponding to small Stokes and finite terminal sedimenting velocity
as in \citet{M87}. In both cases, the resulting expressions for the
eddy-diffusivity tensor greatly simplify: they still involve an
auxiliary problem, but now in the three-dimensional space. Also in this case, in general, only numerical
approaches are capable of obtaining the eddy-diffusivity tensor.
At the perturbative order here considered, gravity does not affect the
diffusion process in the first asymptotics, while it does in the second one.\\
As we mentioned above, the eddy diffusivity is a very old idea in
turbulence \citep[see, e.g.,][]{F95}
which permitted to simulate high Reynolds/P\'eclet number flows.
Motivated by this consideration, we have isolated from both our asymptotics
(vanishing or finite sedimenting velocity) a class of relevant flow fields
from which the eddy diffusivity can be computed analytically.
This is indeed the case for the class of parallel flows.
The aim was to understand if gravity plays to enhance or reduce
the rate of transport. Answering this question might help researchers
involved in closure problems to try
to parameterize the role of gravity in the diffusion process
in terms of relevant properties of the flow field.
Our results are not encouraging
from this point of view: we performed a detailed sensitivity study, from
which it clearly emerges a dependence of the effect played by gravity and inertia
on the details of the carrying flow. A simple, universal, way to
parameterize the role of gravity and inertia
on the diffusion process seems to be
very difficult to achieve.

The paper is organized as follows. In \S~\ref{sec:mult} we sketch the problem under
consideration by recalling the significant equations, and we perform a multiscale
expansion to obtain the eddy-diffusivity tensor and the auxiliary equation.
In \S~\ref{sec:smin} we analyse our small-inertia expansion order by order.
Section \ref{sec:paral} shows explicit results for parallel flows
and namely for a specific case, the well-known Kolmogorov flow (steady or random in time).
Conclusions and perspectives follow in \S~\ref{sec:conc}.
The appendices \S~\ref{ap:det} and \S~\ref{ap:OU} are devoted to show
the mathematical details of the calculation and to recall some basic notions
about the Ornstein--Uhlenbeck process, respectively.

\section{Large-scale dynamics and multiscale expansion} \label{sec:mult}

The model of transport we consider here refers to
point-like inertial particles subject to a constant (gravitational) acceleration
$\bm{g}$ (for the sake of generality,
in $d$ spatial dimensions) and carried by
an \emph{incompressible} velocity field $\bm{u}(\bm{x},t)$.
Beside incompressibility, we suppose that the velocity field is steady or periodic in time
(with period $\mathcal{T}$), and periodic in space with unit cell $\mathbb{B}$ of linear
size $\ell$. It is not a restriction to focus on
velocity fields the average of which vanishes over $\mathbb{B}$:
\begin{equation} \label{labo}
 \int_{\mathbb{B}}\!\ud\bm{x}\,\bm{u}(\bm{x},t)=\bm{0}\;.
\end{equation}
The same technique can be extended to handle the case of a random, homogeneous, and stationary velocity field
with some nontrivial modifications in the rigorous proofs of convergence \citep{AM91}.\\
Neglecting any feedback effect on the transporting fluid and taking only
into account the added-mass effect in a simplified way as in \citet{MA08},
the Lagrangian dynamics reduces to the following set of
stochastic differential equations for particle position ($\bm{\mathcal{X}}(t)$)
and co-velocity ($\bm{\mathcal{V}}(t)$) \citep{MR83,G83}:
\begin{equation} \label{dyn}
 \left\{\begin{array}{l}
  \dot{\bm{\mathcal{X}}}(t)=\bm{\mathcal{V}}(t)+\beta\bm{u}(\bm{\mathcal{X}}(t),t)\\[0.3cm]
  \dot{\bm{\mathcal{V}}}(t)=\displaystyle-\frac{\bm{\mathcal{V}}(t)-(1-\beta)\bm{u}[\bm{\mathcal{X}}(t),t]}{\tau_{\mathrm{S}}}+(1-\beta)\bm{g}+\frac{\sqrt{2\,\kappa}}{\tau_{\mathrm{S}}}\bm{\eta}(t)\;.
 \end{array}\right.
\end{equation}
In (\ref{dyn}) $\bm{\eta}(t)$ is a white-noise process coupled by a constant
diffusivity $\kappa$ \citep{R88}, the pure number $\beta$ is the added-mass factor
$\beta:=3\rho_{\mathrm{f}}/(\rho_{\mathrm{f}}+2\rho_{\mathrm{p}})\in[0,3]$
built from the constant fluid ($\rho_{\mathrm{f}}$) and particle
($\rho_{\mathrm{p}}$) mass densities.
Non-vanishing values of $\beta$, i.e.\ particles not much heavier than the fluid,
induce a discrepancy between the particle velocity,
$\dot{\bm{\mathcal{X}}}(t)$, and co-velocity, $\bm{\mathcal{V}}(t)$ \citep{B03}.
Finally, the Stokes time $\tau_{\mathrm{S}}$ expresses the typical response delay
of particles to flow variations.
For spherical inertial particles of radius $R$, the Stokes time is related to the
kinematic viscosity $\nu$ of the carrier fluid by $\tau_{\mathrm{S}}= R^2/(3\nu\beta)$.

A standard application of stochastic calculus \citep[see, e.g.,][]{C43,V07} yields the Kramers
equation for $\rho(\bm{x},\bm{v},t)$, the phase-space density:
\begin{equation} \label{fp}
 \left\{\frac{\partial}{\partial t}+\frac{\partial}{\partial\bm{x}}\cdot\left(\bm{v}+\beta\bm{u}\right)+\frac{\partial}{\partial\bm{v}}\cdot\left[\frac{(1-\beta)\bm{u}-\bm{v}}{\tau_{\mathrm{S}}}+(1-\beta)\bm{g}\right]-\frac{\kappa}{\tau_{\mathrm{S}}^2}\Delta_{\bm{v}}\right\}\rho=0\;.
\end{equation}
Our aim is to derive the large-scale asymptotics of the marginal density $P$,
\begin{equation} \label{marginal}
 P(\bm{x},t):=\int_{\mathbb{R}^{d}}\!\ud\bm{v}\,\rho(\bm{x},\bm{v},t)\;,
\end{equation}
describing variations of the configuration-space marginal distribution
over spatial and temporal scales $O(x)\approx L\gg\ell$ and $O(t)\approx T\gg\mathcal{T}$, respectively.
The technical tool we employ to extricate this large-scale asymptotics
from the full dynamics is known as \emph{homogenization} or
\emph{multiple-scale expansion} \citep{BLP78,BO78,PS07}.
The idea is to use the scale-separation parametre
\[\epsilon:=\ell/L\ll1\]
to introduce a new set of (``slow'' or ``large-scale'') variables
\begin{equation} \label{variables}
 \left\{\begin{array}{l}
  \bm{X}:=\epsilon\bm{x}\\
  T:=\epsilon^2t\\
  T^{\ddag}:=\epsilon t\;,
 \end{array}\right.
\end{equation}
which we shall treat as independent from the corresponding ``fast'' variables $\bm{x}$ and $t$.
We do not introduce any slow variable $\bm{V}$ associated to the co-velocity $\bm{v}$, since we are
interested only in the asymptotics of the marginal configuration-space distribution (\ref{marginal}).
We define instead the large-scale time variables $T$ and $T^{\ddag}$ to decouple
diffusive from ballistic degrees of freedom \citep{MMV05}. We also hypothesize that the
expression of the initial conditions for (\ref{fp}) in terms of fast and slow variables takes the form
\begin{equation} \label{initial}
 \rho(\bm{x},\bm{X},\bm{v},0,0,0)=\bar{\rho}_{x}(\bm{x})\, \bar{\rho}_{v}(\bm{v})\,\bar{P}(\bm{X})\;,
\end{equation}
where $\bar{\rho}_{x}$ has the same periodicity of the velocity field $\bm{u}$.
The product form (\ref{initial}) is non-restrictive owing to the linearity
of the dynamics (\ref{fp}). This latter is couched by the change of variables
(\ref{variables}),
\[\frac{\partial}{\partial\bm{x}}\mapsto\frac{\partial}{\partial\bm{x}}+\epsilon\frac{\partial}{\partial\bm{X}}\;,\qquad\frac{\partial}{\partial t}\mapsto\frac{\partial}{\partial t}+\epsilon\frac{\partial}{\partial T^{\ddag}}+\epsilon^2\frac{\partial}{\partial T}\;,\qquad\frac{\partial}{\partial\bm{v}}\mapsto\frac{\partial}{\partial\bm{v}}\;,\]
into the form
\begin{equation} \label{fpexp}
 \left(\mathcal{L}^{(0)}+\epsilon\,\mathcal{L}^{(1)}+\epsilon^2\mathcal{L}^{(2)}\right)\rho=0\;,
\end{equation}
with
\begin{subequations}
 \begin{eqnarray} \label{fp:fast}
  \mathcal{L}^{(0)}:=\frac{\partial}{\partial t}+\frac{\partial}{\partial\bm{x}}\cdot\left(\bm{v}+\beta\bm{u}\right)+\frac{\partial}{\partial\bm{v}}\cdot\left[\frac{(1-\beta)\bm{u}-\bm{v}}{\tau_{\mathrm{S}}}+(1-\beta)\bm{g}\right]-\frac{\kappa}{\tau_{\mathrm{S}}^2}\Delta_{\bm{v}}\;,
 \end{eqnarray}
 \begin{eqnarray} \label{fp:mixed}
  \mathcal{L}^{(1)}:=\left(\bm{v}+\beta\bm{u}\right)\cdot\frac{\partial}{\partial\bm{X}}+\frac{\partial}{\partial T^{\ddag}}\;,
 \end{eqnarray}
 \begin{eqnarray} \label{fp:slow}
  \mathcal{L}^{(2)}:=\frac{\partial}{\partial T}\;.
 \end{eqnarray}
\end{subequations}
Our program is then to solve (\ref{fpexp}) in a power-series expansion in the scale-separation parametre,
\begin{subequations}
 \begin{eqnarray} \label{fp:expansion}
  \rho(\bm{x},\bm{X},\bm{v},t,T,T^{\ddag})=\sum_{n=0}^{\infty}\epsilon^n\rho^{(n)}(\bm{x},\bm{X},\bm{v},t,T,T^{\ddag})\;,
 \end{eqnarray}
with initial conditions
 \begin{eqnarray} \label{fp:initial}
  \rho^{(n)}(\bm{x},\bm{X},\bm{v},0,0,0)=\delta_{n0}\bar{\rho}_{x}(\bm{x})\bar{\rho}_{v}(\bm{v})\bar{P}(\bm{X})\;,
 \end{eqnarray}
\end{subequations}
based on the ``unperturbed'' Cauchy problem in the fast variables,
\begin{subequations}
 \begin{eqnarray} \label{fp:unperturbed}
  \mathcal{L}^{(0)}\rho^{(0)}=0\;,
 \end{eqnarray}
 \begin{eqnarray} \label{fp:upinit}
  \rho^{(0)}(\bm{x},\bm{X},\bm{v},0,T,T^{\ddag})=\bar{\rho}_{x}(\bm{x})\bar{\rho}_{v}(\bm{v})P^{(0)}(\bm{X},T,T^{\ddag})\;.
 \end{eqnarray}
\end{subequations}
The final objective is
to attain closed equations for the large-scale dynamics, at the leading order described
by $P^{(0)}$, in the form of solvability conditions for the perturbative hierarchy.
As a result, the details of the small-scale dynamics are summarized in the definition
of renormalized transport coefficients, such as the \emph{effective terminal velocity}
\citep[see also][and references therein]{M87,MA08} and the \emph{eddy diffusivity},
for the advective and diffusive rescalings respectively.\\
For what concerns the fast variables, here the aim is to find square-integrable solutions which enjoy the same space-time periodicity as the flow
field $\bm{u}$, and which decay sufficiently fast at infinity in the co-velocity variable.
Under physically reasonable regularity assumptions
on $\bm{u}$, the results of chapter $3$ of \citet{BLP78} carry over to the operator $\mathcal{L}^{(0)}$.
Its adjoint on the space $\mathbb{L}^2(\mathbb{B}\times\mathbb{R}^{d}\times[0,\mathcal{T}])$,
\begin{eqnarray} \label{adjoint}
 \mathcal{L}^{(0)\dagger}:=-\frac{\partial}{\partial t}-\left(\bm{v}+\beta \bm{u}\right)\cdot\frac{\partial}{\partial\bm{x}}-\left[\frac{(1-\beta)\bm{u}-\bm{v}}{\tau_{\mathrm{S}}}+(1-\beta)\bm{g}\right]\cdot\frac{\partial}{\partial\bm{v}}-\frac{\kappa}{\tau_{\mathrm{S}}^2}\Delta_{\bm{v}}\;,
\end{eqnarray}
has as zero modes only the constants with respect to the fast variables. By the Fredholm alternative,
the existence and uniqueness of the inverse of $\mathcal{L}^{(0)}$ \citep{BLP78,PS07}
is guaranteed by imposing that the kernel of $\mathcal{L}^{(0)\dagger}$ be orthogonal
to the perturbations spawned by (\ref{fp:mixed}) and (\ref{fp:slow}),
or in other words that the integrals of the right-hand sides of (\ref{o1b}) and (\ref{numero}) be vanishing.
This will allow us to determine the dependence on the slow variables.

\subsection{Derivation of the large-scale evolution} \label{sec:Fredholm}

Since the Kramers evolution (\ref{fp}) is measure preserving, it is not restrictive to posit
\begin{equation} \label{Fredholm:npnorm}
 \mea\int_{\mathbb{R}^{d}}\!\ud\bm{X}\,\rho(\bm{x},\bm{X},\bm{v},t,T,T^{\ddag})=1\;.
\end{equation}
Combined with (\ref{fp:initial}), (\ref{Fredholm:npnorm}) allows us to fix the
normalization of the terms of the multiscale expansion to
\begin{equation} \label{Fredholm:ptnorm}
 \mea\int_{\mathbb{R}^{d}}\!\ud\bm{X}\,\rho^{(n)}=\delta_{n0}\;.
\end{equation}
Since the solution of the unperturbed Cauchy problem (\ref{fp:unperturbed}--\ref{fp:upinit}) takes the form
\begin{equation} \label{Fredholm:zero}
 \rho^{(0)}(\bm{x},\bm{X},\bm{v},t,T,T^{\ddag})=p^{(0)}(\bm{x},\bm{v},t)P^{(0)}(\bm{X},T,T^{\ddag})\;,
\end{equation}
with
\begin{equation} \label{ref}
 \mathcal{L}^{(0)}p^{(0)}(\bm{x},\bm{v},t)=0\;,
\end{equation}
from (\ref{Fredholm:ptnorm}) we obtain
\[\mea\,p^{(0)}(\bm{x},\bm{v},t)=1=\!\int_{\mathbb{R}^{d}}\!\ud\bm{X}\,P^{(0)}(\bm{X},T,T^{\ddag})\;.\]
We will show below that this condition is mathematically self-consistent because
the solvability conditions yield for $P$ a measure-preserving
evolution law. Physically, this is desirable as it corresponds to the statement that,
when focusing on the sole large scales of the problem, $P$ can be interpreted itself
as a mass density. It is also a necessary condition to correctly reproduce the well-known
small-inertia limit for the eddy diffusivity of tracer particles.

\subsubsection{Solvability condition for $\rho^{(1)}$: large-scale ballistic motion}

At $O(\epsilon^1)$ we have:
\begin{equation} \label{o1b}
 \mathcal{L}^{(0)}\rho^{(1)}=-\mathcal{L}^{(1)}\rho^{(0)}=-\left[\left(\bm{v}+\beta\bm{u}\right)\cdot\frac{\partial}{\partial\bm{X}}+\frac{\partial}{\partial T^{\ddag}}\right]\rho^{(0)}\;.
\end{equation}
By the Fredholm alternative the problem admits a unique solution if and only if
\begin{equation} \label{numerata}
 \mea\,\mathcal{L}^{(1)}\rho^{(0)}=-\!\mea\,\mathcal{L}^{(0)}\rho^{(1)}=0\;.
\end{equation}
Upon inserting (\ref{Fredholm:zero}) into (\ref{numerata}) we get the solvability condition
\begin{equation} \label{bal}
 \frac{\partial}{\partial T^{\ddag}}P^{(0)}(\bm{X},T,T^{\ddag})+\bm{w}\cdot\frac{\partial}{\partial\bm{X}}P^{(0)}(\bm{X},T,T^{\ddag})=0\;,
\end{equation}
with $\bm{w}$ the \emph{terminal velocity} \citep{M87,MA08} specified by
\begin{equation} \label{vt}
  \bm{w}:=\!\mea\,\left[\bm{v}+\beta\bm{u}(\bm{x},t)\right]p^{(0)}(\bm{x},\bm{v},t)\;.
\end{equation}
The physical interpretation of (\ref{bal}) is to associate the terminal velocity
to ballistic large-scale motion. The solution of (\ref{bal}) requires $P^{(0)}$
to satisfy
\[ P^{(0)}(\bm{X},T,T^{\ddag})=P^{(0)}(\bm{X}-\bm{w}\,T^{\ddag},T,0)\]
without fixing the functional dependence with respect to the first two arguments.
As in \citet{MMV05} we can leverage this fact to find the function
$P_*^{(0)}(\bm{X},T)$ satisfying the Fredholm alternative condition at the second order
in $\epsilon$. We can then reconstruct the full solution by the Galilean transformation
\[P^{(0)}(\bm{X},T,T^{\ddag}):=P_*^{(0)}(\bm{X}-\bm{w}\,T^{\ddag},T)\;.\]

\subsubsection{Solution of the equation for the leading-order correction $\rho^{(1)}$}

Exploiting (\ref{bal}), the first-order equation (\ref{o1b}) reduces to:
\begin{equation} \label{Fredholm:one}
 \mathcal{L}^{(0)}\rho^{(1)}=-p^{(0)}\left(\bm{v}+\beta\bm{u}-\bm{w}\right)\cdot\frac{\partial P^{(0)}}{\partial\bm{X}}\;.
\end{equation}
The initial condition (\ref{fp:initial}) rules out
the presence in $\rho^{(1)}$ of any solution of the associated homogeneous
problem. We are therefore entitled to couch the solution into the form
\begin{equation} \label{adhoc}
 \rho^{(1)}(\bm{x},\bm{X},\bm{v},t,T,T^{\ddag})=\bm{p}^{(1)}(\bm{x},\bm{v},t)\cdot\frac{\partial P^{(0)}}{\partial\bm{X}}(\bm{X},T,T^{\ddag})\;.
 \end{equation}
with $\bm{p}^{(1)}$ a $d$-dimensional vector field satisfying
\begin{subequations}
 \begin{eqnarray} \label{ea}
  \mathcal{L}^{(0)}\bm{p}^{(1)}(\bm{x},\bm{v},t)=-\left[\bm{v}+\beta\bm{u}(\bm{x},t)-\bm{w}\right]p^{(0)}(\bm{x},\bm{v},t)\;,
 \end{eqnarray}
 \begin{eqnarray} \label{ci}
  \bm{p}^{(1)}(\bm{x},\bm{v},0)=\bm{0}\;.
 \end{eqnarray}
\end{subequations}
Equation (\ref{ea}) is usually called ``auxiliary equation'' or ``cell problem''.
The advantage of (\ref{adhoc}) is that it satisfies at glance the normalization
condition (\ref{Fredholm:ptnorm}) with $n=1$, in consequence of the gradient form in the slow variable $\bm{X}$.

\subsubsection{Solvability condition for $\rho^{(2)}$: large-scale diffusive motion}

Large-scale diffusive effects set in only at order $O(\epsilon^2)$. Namely,
\begin{equation} \label{numero}
 \mathcal{L}^{(0)}\rho^{(2)}=-\mathcal{L}^{(1)}\rho^{(1)}-\mathcal{L}^{(2)}\rho^{(0)}=-\left[\left(\bm{v}+\beta\bm{u}\right)\cdot\frac{\partial}{\partial\bm{X}}+\frac{\partial}{\partial T^{\ddag}}\right]\rho^{(1)}-\frac{\partial}{\partial T}\rho^{(0)}
\end{equation}
admits a unique solution if and only if
\begin{equation} \label{Fredholm:cs2}
 \mea\,\left(\mathcal{L}^{(1)}\rho^{(1)}+\mathcal{L}^{(2)}\rho^{(0)}\right)=-\mea\,\mathcal{L}^{(0)}\rho^{(2)}=0\;.
\end{equation}
Upon inserting on the right-hand side of (\ref{Fredholm:cs2}) the explicit expressions
(\ref{fp:mixed}) of $\mathcal{L}^{(1)}$ and (\ref{adhoc}) of $\rho^{(1)}$, after straightforward algebra we obtain
\begin{equation} \label{cs2}
  \frac{\partial}{\partial T}P(\bm{X},T,T^{\ddag})=K_{\mu\nu}\frac{\partial^2}{\partial X_{\mu}\partial X_{\nu}}P(\bm{X},T,T^{\ddag})\;,
\end{equation}
where we have introduced the \emph{eddy-diffusivity} tensor
\begin{equation} \label{de}
 K_{\mu\nu}:=-\mathcal{S}\!\mea\,\left[v+\beta u(\bm{x},t)-w\right]_{\mu}p^{(1)}_{\nu}(\bm{x},\bm{v},t)\;,
\end{equation}
with $\mathcal{S}$ the trace-preserving index symmetrization operator: $\mathcal{S}\mathsf{A}_{\mu\nu}:=(\mathsf{A}_{\mu\nu}+\mathsf{A}_{\nu\mu})/2$.\\
However, finding explicit expressions for (\ref{de}) requires the knowledge of the terminal velocity $\bm{w}$ (\ref{vt}) and of the vector field $\bm{p}^{(1)}$ (\ref{ea}), and thus also of the density $p^{(0)}$ (\ref{ref}). These two latter partial differential equations are defined in a $(2d+1)$-dimensional space according to the definition of $\mathcal{L}^{(0)}$ (\ref{fp:fast}), and are thus prohibitive to solve also numerically. In \S~\ref{sec:smin} we will turn to the derivation of systematic approximations for these quantities in the limit of small inertia, which allows us to decouple the co-velocity degree of freedom and therefore to recast the problem in terms of ``innocent'' forced advection--diffusion equations, which are $(d+1)$-dimensional and can be solved relatively easily for specific flows (\S~\ref{sec:paral}) or in general from the numerical point of view.\\
In order to prove the diffusive nature of the large-scale motion described by (\ref{cs2}),
it remains to show that the eddy-diffusivity tensor (\ref{de}) is positive definite.
The proof proceeds along the same lines as in \citet{PS05}, therefore we only briefly recall its outline.
By (\ref{vt}) and the Fredholm alternative, the auxiliary equation
\begin{equation} \label{PS2}
 \mathcal{L}^{(0)\dagger}\bm{\phi}(\bm{x},\bm{v},t)=\bm{v}+\beta\bm{u}(\bm{x},t)-\bm{w}
\end{equation}
admits a unique solution. We can therefore exploit it to write (\ref{de}) as a
matrix element of $\mathcal{L}^{(0)\dagger}$. An integration by parts and
the use of (\ref{ea}) and of (\ref{PS2}) again, finally give
\[K_{\mu\nu}=\mathcal{S}\!\mea\,p^{(0)}(\bm{x},\bm{v},t)\left[v+\beta u(\bm{x},t)-w\right]_{\mu}\mathcal{L}^{(0)\dagger-1}\left[v+\beta u(\bm{x},t)-w\right]_{\nu}\;,\]
which is essentially a weighted average of the projection of a positive-definite
operator onto the complement of its kernel.

The solution $P_*^{(0)}$ of (\ref{cs2}) satisfying the initial condition
\[P_*^{(0)}(\bm{X},0)=\bar{P}(\bm{X})\]
yields the large-scale asymptotics we set out to derive
\begin{equation} \label{Fredholm:solution}
 P(\bm{x},t)=P_*^{(0)}(\epsilon\bm{x}-\epsilon\bm{w}t,\epsilon^2t)+O(\epsilon)\;.
\end{equation}

\section{Small-inertia expansion} \label{sec:smin}

It is expedient to define the $(d+1)$-dimensional vector fields $\bm{\psi}$ and $\bm{W}$ with components%
\footnote{Notice that $\psi_0$ thus turns out to have dimensions (length$^{-d}\times$velocity$^{-d}$)
 different from those of the remaining components (length$^{1-d}\times$velocity$^{-d}$).
 One could easily fix this point by setting $\psi_0:=\ell p^{(0)}$,
 but we prefer to avoid such an operation for the sake of notational simplicity.}
\[\left\{\begin{array}{lll}
 \psi_{\mu}:=p^{(0)}&\textrm{and }W_{\mu}:=0\;,&\textrm{for }\mu=0\\[0.2cm]
 \psi_{\mu}:=p^{(1)}_{\mu}&\textrm{and }W_{\mu}:=v_{\mu}+\beta u_{\mu}(\bm{x},t)-w_{\mu}\;,&\textrm{for }1\leq\mu\leq d\;.
\end{array}\right.\]
From (\ref{ref}) and (\ref{ea}) it follows immediately that:
\begin{subequations}
 \begin{eqnarray} \label{smin:eq}
  \mathcal{L}^{(0)}\psi_{\mu}(\bm{x},\bm{v},t)=-W_{\mu}(\bm{x},\bm{v},t)\psi_0(\bm{x},\bm{v},t)\;,
 \end{eqnarray}
 \begin{eqnarray} \label{smin:init}
  \psi_{\mu}(\bm{x},\bm{v},0)=\delta_{\mu0}\bar{\rho}_x(\bm{x})\bar{\rho}_v(\bm{v})\;.
 \end{eqnarray}
\end{subequations}
The problem of deriving systematic approximations for the auxiliary fields $p^{(0)}$,
$\bm{p}^{(1)}$ reduces to that of solving the spectral properties
of $\mathcal{L}^{(0)}$ in some suitable limit.
The \emph{small-inertia} limit recovering the over-damped Fokker--Planck
from Kramers' dynamics \citep[see, e.g.,][]{G85,R89,V07} is a natural candidate.
To set the scene, let us denote by $U$ the typical magnitude of
the advecting velocity field ($U$ may be defined e.g.\ as the space-time average of $||\bm{u}||(\bm{x},t)$
over $[0,\mathcal{T}]\times\mathbb{B}$). From $U$, the unit cell
linear size $\ell$, the Stokes time $\tau_{\mathrm{S}}$, the Brownian diffusivity $\kappa$
and the gravitational acceleration $g$, we can construct the hydrodynamical numbers
\[\textrm{Stokes: }\mathrm{St}:=\frac{\tau_{\mathrm{S}}U}{\ell}\;,\qquad\textrm{Froude: }\mathrm{Fr}:=\frac{U}{\sqrt{g\ell}}\;,\qquad\textrm{P\'eclet: }\mathrm{Pe}:=\frac{\ell U}{\kappa}\;.\]
The small-inertia limit corresponds to let $\mathrm{St}$ tend to zero
for arbitrary but finite values of $\mathrm{Fr}$ and $\mathrm{Pe}$. In such a limit the Stokes time sets
the shortest time scale in the problem. The corresponding typical co-velocity variation
is diffusion dominated. The root mean square along any Cartesian axis,
\[v_{\tau}=\sqrt{\frac{2\kappa}{\tau_{\mathrm{S}}}}=\sqrt{\frac{2}{\mathrm{St}\,\mathrm{Pe}}}\,U\;,\]
implies that co-velocities become fast variables in units of $U$ for time scales of the
order of $\tau_{\mathrm{S}}$. We therefore attune the functional dependence on $\mathrm{St}$ to the
description of slow degrees of freedom by working with rescaled co-velocities,
\begin{equation} \label{smin:vel}
 \bm{v}^{\prime}:=\sqrt{\frac{\mathrm{St}\,\mathrm{Pe}}{2}}\,\bm{v}
\end{equation}
and by decoupling the time dependence of solutions of (\ref{smin:eq}) into a fast ($t^{\ddag}$)
and a slow ($t^{\prime}$) component,
so as to deal now with a new vector field $\bm{\psi}^*(\bm{x},\bm{v}^{\prime},t^{\prime},t^{\ddag})$,
which is trivially related to its original counterpart $\bm{\psi}$ through the multiplication by the Jacobian.

Upon inserting
\[\frac{\partial}{\partial t}=\frac{1}{\mathrm{St}}\frac{\partial}{\partial t^{\ddag}}+\frac{\partial}{\partial t^{\prime}}\]
into (\ref{smin:eq}), the variables (\ref{smin:vel}) yield for $\mathcal{L}^{(0)}$ the expansion
\[\mathcal{L}^{(0)}=\sum_{n=0}^{3}\mathrm{St}^{n/2-1}\mathcal{L}^{(0:n)}\]
where (omitting from now on the prime over-script to simplify the notation):
\begin{subequations}
 \begin{eqnarray} \label{smin:OU}
  \mathcal{L}^{(0:0)}:=\frac{\partial}{\partial t^{\ddag}}-\frac{\sigma_{\ell}^2}{\tau_{\ell}}\Delta_{\bm{v}}-\frac{1}{\tau_{\ell}}\frac{\partial}{\partial\bm{v}}\cdot\bm{v}\;,
 \end{eqnarray}
 \begin{eqnarray} \label{smin:mixed}
  \mathcal{L}^{(0:1)}:=\frac{1-\beta}{\tau_{\ell}}\sqrt{\frac{\mathrm{Pe}}{2}}\,\bm{u}\cdot\frac{\partial}{\partial\bm{v}}+\sqrt{\frac{2}{\mathrm{Pe}}}\,\bm{v}\cdot\frac{\partial}{\partial\bm{x}}\;,
 \end{eqnarray}
 \begin{eqnarray} \label{smin:slow}
  \mathcal{L}^{(0:2)}:=\frac{\partial}{\partial t}+\beta\bm{u}\cdot\frac{\partial}{\partial\bm{x}}\;,
 \end{eqnarray}
 \begin{eqnarray} \label{smin:neglected}
  \mathcal{L}^{(0:3)}:=\frac{\sigma_{\ell}(1-\beta)}{\tau_{\ell}\mathrm{Fr}^2}\sqrt{\mathrm{Pe}}\,\bm{G}\cdot\frac{\partial}{\partial\bm{v}}\;.
 \end{eqnarray}
\end{subequations}
Here and in what follows, the digit after the colon in the superscripts represents the order of the expansion in $\mathrm{St}$,
keeping in mind that the latter is in half-integer powers and for some quantities might not start from the zeroth order;
this notation was used in order to distinguish such an expansion from the multiscale one shown in the previous section.
We defined in (\ref{smin:OU})
\[\sigma_{\ell}^2:=\frac{U^2}{2}\;,\qquad\tau_{\ell}:=\frac{\ell}{U}\]
to identify the variance of the co-velocity steady-state Maxwell--Boltzmann distribution
(see \S~\ref{sec:MA08p} below) and the advective time scale, and in (\ref{smin:neglected})
\[\bm{G}:=\frac{\bm{g}}{||\bm{g}||}\]
to exhibit the Froude number dependence.\\
Two remarks are here in order. First, the effect of the rescaling (\ref{smin:vel}) is to
balance drift and diffusion terms in the co-velocity dynamics for vanishing $\mathrm{St}$.
The solution of the auxiliary problem is thus turned into a multiscale expansion in powers of
$\mathrm{St}$ based on the well-known Ornstein--Uhlenbeck operator (\ref{smin:OU})
(see \S~\ref{ap:OU} for further details).
The kernel of $\mathcal{L}^{(0:0)}$ defines a Maxwell--Boltzmann co-velocity distribution
to which any integrable initial density relaxes for time scales much larger than the
Stokes time $\tau_{\mathrm{S}}$. Second, the functional dependence on the faster time $t^{\ddag}$ is only of conceptual
importance. It only describes the exponentially fast equilibration process of the
marginal co-velocity distribution to the aforementioned Maxwell--Boltzmann steady state,
and can therefore be neglected in the asymptotics we are interested in.
The upshot is that we can restrict the attention to the $t^{\ddag}$-asymptotic
form of the power-series expansion
\[\bm{\psi}^*(\bm{x},\bm{v},t,t^{\ddag})\xrightarrow{t^{\ddag}\nearrow\infty}\sum_{n=0}^{\infty}\mathrm{St}^{n/2}\bm{\psi}^{(:n)}(\bm{x},\bm{v},t)\;,\]
and take into account the initial condition (\ref{smin:init}) only to the extent
that it affects the time dependence upon the slower time variable.
With this proviso, the expansion of the terminal velocity (\ref{vt}) in powers
of $\mathrm{St}$ yields
\begin{equation} \label{smin:vt}
 \bm{w}=\frac{\bm{w}^{(:-1)}}{\mathrm{St}^{1/2}}+\sum_{n=0}^{\infty}\mathrm{St}^{n/2}\bm{w}^{(:n)}\;,
\end{equation}
where
\begin{equation} \label{smin:vt_coe}
 \bm{w}^{(:n)}=\left\{\begin{array}{ll}
  \displaystyle\mea\,\sqrt{\frac{2}{\mathrm{Pe}}}\,\bm{v}\psi^{(:0)}_0\hspace{0.3cm}&\hspace{0.3cm}n=-1\\[0.3cm]
  \displaystyle\mea\,\left(\sqrt{\frac{2}{\mathrm{Pe}}}\,\bm{v}\psi^{(:n+1)}_0+\beta\bm{u}\psi^{(:n)}_0\right)\hspace{0.3cm}&\hspace{0.3cm}n\geq0\;,
 \end{array}\right.
\end{equation}
so that the auxiliary equation (\ref{ea}) spawns a perturbative hierarchy
the first terms whereof are (no sum on $\mu$ is implied here):
\begin{itemize}
 \item[$O(\mathrm{St}^{-1})$]:
  \begin{equation} \label{prima}
   \mathcal{L}^{(0:0)}\psi^{(:0)}_{\mu}=0\;;
  \end{equation}
 \item[$O(\mathrm{St}^{-1/2})$]:
  \begin{equation} \label{seconda}
   \mathcal{L}^{(0:0)}\psi^{(:1)}_{\mu}=-\mathcal{L}^{(0:1)}\psi^{(:0)}_{\mu}-(1-\delta_{\mu0})\left(\sqrt{\frac{2}{\mathrm{Pe}}}\,v_{\mu}-w^{(:-1)}_{\mu}\right)\psi^{(:0)}_0\;;
  \end{equation}
 \item[$O(\mathrm{St}^0)$]:
  \begin{eqnarray} \label{terza}
   \lefteqn{\mathcal{L}^{(0:0)}\psi^{(:2)}_{\mu}=-\sum_{n=0}^1\mathcal{L}^{(0:2-n)}\psi^{(:n)}_{\mu}}\nonumber\\
   &&-(1-\delta_{\mu0})\left[\left(\sqrt{\frac{2}{\mathrm{Pe}}}\,v_{\mu}-w^{(:-1)}_{\mu}\right)\psi^{(:1)}_0+\left(\beta u_{\mu}-w^{(:0)}_{\mu}\right)\psi^{(:0)}_0\right]\;;
  \end{eqnarray}
 \item[$O(\mathrm{St}^{1/2})$]:
  \begin{eqnarray} \label{quarta}
   \lefteqn{\mathcal{L}^{(0:0)}\psi^{(:3)}_{\mu}=-\sum_{n=0}^2\mathcal{L}^{(0:3-n)}\psi^{(:n)}_{\mu}}\nonumber\\
   &&-(1-\delta_{\mu0})\left[\left(\sqrt{\frac{2}{\mathrm{Pe}}}\,v_{\mu}-w^{(:-1)}_{\mu}\right)\psi^{(:2)}_0+\left(\beta u_{\mu}-w^{(:0)}_{\mu}\right)\psi^{(:1)}_0-w^{(:1)}_{\mu}\psi^{(:0)}_0\right]\;.
 \end{eqnarray}
\end{itemize}
We solved (\ref{prima})--(\ref{quarta}) by first deriving
the coefficients of the expansion of $\psi_0\equiv p^{(0)}$ (the phase-space density)
to thereafter use them for the evaluation of the remaining components of the
auxiliary vector field $\bm{\psi}_{\ge1}\equiv\bm{p}^{(1)}$. The details of the calculation can be found in \S~\ref{ap:det},
more precisely in \S~\ref{sec:MA08p} for the first step and in \S~\ref{sec:psi} for the second one.

\subsection{Expression of the eddy diffusivity} \label{sec:kappa}

Within third-order accuracy, the co-velocity dependence of both the phase-space density $p^{(0)}$
and the auxiliary vector field $\bm{p}^{(1)}$ is parity defined (see \S~\ref{ap:det}). In consequence, the Stokes-number
expansion of the eddy diffusivity (\ref{de}) tensor reduces to:
\begin{eqnarray} \label{kappamunu}
 \lefteqn{K_{\mu\nu}=-\mathcal{S}\!\mea\,\left(\sqrt{\frac{2}{\mathrm{Pe}}}\,v_{\mu}\psi^{(:1)}_{\nu}+\beta u_{\mu}\psi^{(:0)}_{\nu}\right)}\nonumber\\
 &&-\mathrm{St}\,\mathcal{S}\!\mea\,\left(\sqrt{\frac{2}{\mathrm{Pe}}}\,v_{\mu}\psi^{(:3)}_{\nu}+\beta u_{\mu}\psi^{(:2)}_{\nu}\right)+O(\mathrm{St}^2)\;.
\end{eqnarray}
Inserting the terms of the expansion of $\bm{p}^{(1)}$ leads to further simplifications (see \S~\ref{sec:eda}).

\subsubsection{Evaluation of the $O(\mathrm{St}^0)$ term}
The tracer limit of the particle eddy diffusivity (first line of (\ref{kappamunu})) reduces to
\begin{equation} \label{k0}
  K^{(:0)}_{\mu\nu}=\frac{2\tau_{\ell}\sigma_{\ell}^2}{\mathrm{Pe}}\delta_{\mu\nu}-\mathcal{S}\!\mear\,u_{\mu}(\bm{x},t)\xi^{(1:0)}_{\nu}(\bm{x},t)\;,
\end{equation}
with the auxiliary vector $\bm{\xi}^{(1:0)}$ obeying an advection--diffusion equation with zero initial condition:
\begin{subequations}
 \begin{eqnarray} \label{csi0BIS}
  \left[\frac{\partial}{\partial t}+\bm{u}(\bm{x},t)\cdot\frac{\partial}{\partial\bm{x}}-\kappa\Delta_{\bm{x}}\right]\bm{\xi}^{(1:0)}(\bm{x},t)=-\frac{\bm{u}(\bm{x},t)}{\ell^d}\;,
 \end{eqnarray}
 \begin{eqnarray} \label{zic}
  \bm{\xi}^{(1:0)}(\bm{x},0)=\bm{0}\;.
 \end{eqnarray}
\end{subequations}
An alternative rewrite of (\ref{k0}),
\begin{equation} \label{kappa:0}
 K^{(:0)}_{\mu\nu}=\kappa\left[\delta_{\mu\nu}+\!\mear\,\ell^d\frac{\partial\xi^{(1:0)}_{\mu}}{\partial x_{\alpha_1}}(\bm{x},t)\frac{\partial\xi^{(1:0)}_{\nu}}{\partial x_{\alpha_1}}(\bm{x},t)\right]\;,
\end{equation}
shows that this quantity is positive definite and recovers the result of \citet{BCVV95}.

\subsubsection{Evaluation of the $O(\mathrm{St}^1)$ term}
The leading-order correction in (\ref{kappamunu}) turns out to be
\begin{equation} \label{kappa:2}
 K^{(:2)}_{\mu\nu}=-\mathcal{S}\!\mear\,\left[u_{\nu}(\bm{x},t)\xi^{(1:2)}_{\mu}(\bm{x},t)-\frac{\tau_{\ell}(1-\beta)}{\ell^d}u_{\mu}(\bm{x},t)u_{\nu}(\bm{x},t)\right]\;,
\end{equation}
where the auxiliary vector $\bm{\xi}^{(1:2)}$ satisfies another (more complicated) advection--diffusion equation with vanishing initial condition:
\begin{subequations}
 \begin{eqnarray} \label{psi:cc12BIS}
  \lefteqn{\left(\frac{\partial}{\partial t}+\bm{u}\cdot\frac{\partial}{\partial\bm{x}}-\kappa\Delta_{\bm{x}}\right)\xi^{(1:2)}_{\mu}=}\nonumber\\
  &&-\tau_{\ell}(1-\beta)g_{\alpha_1}\frac{\partial}{\partial x_{\alpha_1}}\xi^{(1:0)}_{\mu}+\tau_{\ell}^2\frac{\partial}{\partial x_{\alpha_1}}\left(\sigma_{\ell}\sqrt{\frac{2}{\mathrm{Pe}}}\,\mathcal{L}^{(0:2)}\mathcal{M}_{\alpha_1}+\kappa\frac{\partial}{\partial x_{\alpha_2}}\mathcal{M}_{\alpha_1}\mathcal{M}_{\alpha_2}\right)\xi^{(1:0)}_{\mu}\nonumber\\
  &&+\kappa\frac{\partial}{\partial x_{\mu}}\xi^{(0:2)}+\frac{\tau_{\ell}(1-\beta)}{\ell^d}u_{\alpha_1}\frac{\partial}{\partial x_{\alpha_1}}u_{\mu}-\frac{\tau_{\ell}(1-\beta)\kappa}{2\ell^d}\Delta_{\bm{x}}u_{\mu}\;,
 \end{eqnarray}
 \begin{eqnarray} \label{vic}
  \bm{\xi}^{(1:2)}(\bm{x},0)=0\;.
 \end{eqnarray}
\end{subequations}
Here, $\mathcal{L}^{(0:2)}$ was defined in (\ref{smin:slow}) and
\begin{equation} \label{MA08p:diffBIS}
 \bm{\mathcal{M}}:=\frac{1-\beta}{\tau_{\ell}\sigma_{\ell}}\sqrt{\frac{\mathrm{Pe}}{2}}\,\bm{u}(\bm{x},t)-\sigma_{\ell}\sqrt{\frac{2}{\mathrm{Pe}}}\,\frac{\partial}{\partial\bm{x}}\;.
\end{equation}

\subsubsection{Perturbative expression of the eddy-diffusivity tensor }
Gleaning (\ref{kappa:0}) and (\ref{kappa:2}), we get the final
result for the Stokes-number expansion of the eddy-diffusivity tensor (\ref{de}):
\begin{eqnarray} \label{kappa:final}
 \lefteqn{K_{\mu\nu}=\kappa\left(\delta_{\mu\nu}+\!\mear\,\ell^d\frac{\partial\xi^{(1:0)}_{\mu}}{\partial x_{\alpha_1}}\frac{\partial\xi^{(1:0)}_{\nu}}{\partial x_{\alpha_1}}\right)}\nonumber\\
 &&+\mathrm{St}\!\mear\,\left[\frac{\tau_{\ell}(1-\beta)}{\ell^d}\,u_{\mu}u_{\nu}-\frac{1}{2}\left(u_{\mu}\xi^{(1:2)}_{\nu}+u_{\nu}\xi^{(1:2)}_{\mu}\right)\right]+O(\mathrm{St}^2)\;,
\end{eqnarray}
with $\bm{\xi}^{(1:0)}$ and $\bm{\xi}^{(1:2)}$ given by (\ref{csi0BIS}) and (\ref{psi:cc12BIS}), respectively.
An important physical implication of (\ref{kappa:final}) is that, within $O(\mathrm{St})$-accuracy,
the eddy diffusivity depends only implicitly on gravity through (\ref{psi:cc12BIS}).\\
A more compact rewriting is
\[K_{\mu\nu}=\frac{U\ell}{\mathrm{Pe}}\delta_{\mu\nu}+\!\mear\,\left[-\mathcal{S}u_{\mu}\xi^{(1)}_{\nu}+\mathrm{St}(1-\beta)u_{\mu}u_{\nu}\right]+O(\mathrm{St}^2)\;,\]
with the auxiliary vector $\bm{\xi}^{(1)}$ to be expanded in a power series in $\mathrm{St}$,
whose zeroth and first orders are $\bm{\xi}^{(1:0)}$ and $\bm{\xi}^{(1:2)}$.

\subsection{Expansion at small inertia and finite terminal velocity} \label{sec:Maxey}

The procedure exposed up to now assumes finite $\mathrm{Fr}$, i.e.\ finite gravity effects,
which in the limit of small $\mathrm{St}$ makes $\bm{w}$ infinitesimal.
An alternative point of view \citep{M87,MFS07} consists in assuming the terminal velocity as a finite parametre:
we will investigate this in the present subsection, denoting with a tilde the quantities which undergo some change.\\
A limit of physical interest is thus that of vanishing $\mathrm{St}$ for large values of $\mathrm{Fr}$, so that
\begin{equation} \label{Maxey:limit}
 S_F:=\frac{\mathrm{St}}{\mathrm{Fr}^2}\equiv\frac{g\tau_{\mathrm{S}}}{U}=O(1)\;.
\end{equation}
The limit implies a reshuffling of the hierarchy (\ref{prima})--(\ref{quarta}) since
the operator $\mathcal{L}^{(0:3)}$ (\ref{smin:neglected}) becomes same order as
(\ref{smin:mixed}), so that
\[\mathcal{L}^{(0:1)}\mapsto\tilde{\mathcal{L}}^{(0:1)}:=\frac{1-\beta}{\tau_{\ell}}\sqrt{\frac{\mathrm{Pe}}{2}}\,\left[\bm{u}(\bm{x},t)+S_FU\bm{G}\right]\cdot\frac{\partial}{\partial\bm{v}}+\sqrt{\frac{2}{\mathrm{Pe}}}\,\bm{v}\cdot\frac{\partial}{\partial\bm{x}}\;,\]
whilst
\[\mathcal{L}^{(0:3)}\mapsto\tilde{\mathcal{L}}^{(0:3)}=0\;.\]
The redefinitions affect both the phase-space density $p^{(0)}$ and the auxiliary field $\bm{p}^{(1)}$
(see \S~\ref{sec:app}). Plugging these results into the $\mathrm{St}$-expansion of the eddy diffusivity, we obtain
\begin{equation} \label{kappatilde}
 K_{\mu\nu}\mapsto\tilde{K}_{\mu\nu}=\kappa\left[\delta_{\mu\nu}+\!\mear\,\ell^d\frac{\partial\tilde{\xi}^{(1:0)}_{\mu}}{\partial x_{\alpha_1}}(\bm{x},t)\frac{\partial\tilde{\xi}^{(1:0)}_{\nu}}{\partial x_{\alpha_1}}(\bm{x},t)\right]+O(\mathrm{St})\;,
\end{equation}
which now exhibits an implicit dependence upon gravity already at $O\left(\mathrm{St}^0\right)$
(for this reason here we do not write down the results for $O\left(\mathrm{St}^1\right)$)
through the new following advection--diffusion equation, again supplemented with zero initial condition:
\begin{subequations}
 \begin{eqnarray} \label{csitilde}
  \left\{\frac{\partial}{\partial t}+\left[\bm{u}(\bm{x},t)+(1-\beta)S_FU\bm{G}\right]\cdot\frac{\partial}{\partial\bm{x}}-\kappa\Delta_{\bm{x}}\right\}\tilde{\bm{\xi}}^{(1:0)}(\bm{x},t)=-\frac{\bm{u}(\bm{x},t)}{\ell^d}\;,
 \end{eqnarray}
 \begin{eqnarray} \label{cin}
  \tilde{\bm{\xi}}^{(1:0)}(\bm{x},0)=0\;.
 \end{eqnarray}
\end{subequations}

\section{Specific examples: parallel flows and Kolmogorov flow} \label{sec:paral}

A flow is defined \emph{parallel} if it is always and everywhere in the same direction (e.g.\ along $x_1$),
i.e.\ if the velocity field can be written as $u_{\mu}(\bm{x},t)=\delta_{\mu1}u(\bm{x},t)$.
The incompressibility constraint of zero divergence then requires $u$ to be independent from $x_1$, i.e.
\[u_{\mu}(\bm{x},t)=\delta_{\mu1}u(x_2,\ldots,x_d,t)\;.\]
(Further simplifications can be obtained for specific instances of flow,
e.g.\ if $u_{\mu}(\bm{x},t)=\delta_{\mu1}u(x_d,t)$,
only dependent on one spatial direction along which we orient the $x_d$ axis,
as is the case for the well-known Kolmogorov flow.)

Equations (\ref{csi0BIS}) and (\ref{psi:cc12BIS}),
coupled with vanishing initial conditions for $\bm{\xi}^{(1:0)}$ and $\bm{\xi}^{(1:2)}$,
make $\xi^{(1:0)}_1(\bm{x},t)$ and $\xi^{(1:2)}_1(\bm{x},t)$ the only non-zero vectorial components,
as they are the only ones forced. Moreover, this forcing is independent of $x_1$,
and thus the same independence also holds for $\xi^{(1:0)}_1(x_2,\ldots,x_d,t)$ and $\xi^{(1:2)}_1(x_2,\ldots,x_d,t)$.
This means that the advective terms of the type $\bm{u}\cdot\partial/\partial\bm{x}$ always vanish,
and that the aforementioned equations, which turn out to be heavily simplified,
can easily be solved in Fourier space thanks to the absence of any convolution.
We denote the Fourier transformed by a hat
(for the sake of simplicity, the cell size is assumed as $\ell$ in every spatial dimension),
\[\hat{f}(\bm{n}_{\bm{q}},n_{\omega})=\ell^{-d}\mathcal{T}^{-1}\!\int_0^{\mathcal{T}}\ud t\int_{\mathbb{B}}\ud\bm{x}\,\ue^{-\ui(\bm{q}\cdot\bm{x}+\omega t)}f(\bm{x},t)\;,\]
with $\bm{q}:=2\pi\bm{n}_{\bm{q}}/\ell$ and $\omega:=2\pi n_{\omega}/\mathcal{T}$, so that \[f(\bm{x},t)=\sum_{\bm{n}_{\bm{q}}\in\mathbb{Z}^d}\sum_{n_{\omega}\in\mathbb{Z}}\ue^{\ui(\bm{q}\cdot\bm{x}+\omega t)}\hat{f}(\bm{n}_{\bm{q}},n_{\omega})\;.\]
Therefore, we have:
\begin{equation} \label{csi0fourier}
 \left(\frac{\partial}{\partial t}-\kappa\Delta_{\bm{x}}\right)\xi_{\mu}^{(1:0)}=-\frac{\delta_{\mu1}}{\ell^d}u\Longrightarrow\hat{\xi}_{\mu}^{(1:0)}(\bm{n}_{\bm{q}},n_{\omega})=-\frac{\delta_{\mu1}}{\ell^d}\frac{\hat{u}(\bm{n}_{\bm{q}},n_{\omega})}{\ui\omega+\kappa q^2}\;,
\end{equation}
\begin{eqnarray*} \label{psi:cc12fourier}
 &&\left(\frac{\partial}{\partial t}-\kappa\Delta_{\bm{x}}\right)\xi_{\mu}^{(1:2)}=\left[\kappa^2\tau_{\ell}\Delta_{\bm{x}}^2-\kappa\tau_{\ell}\frac{\partial}{\partial t}\Delta_{\bm{x}}-\tau_{\ell}(1-\beta)\bm{g}\cdot\frac{\partial}{\partial\bm{x}}\right]\xi^{(1:0)}_{\mu}-\frac{\tau_{\ell}(1-\beta)\kappa}{2\ell^d}\Delta_{\bm{x}}u_{\mu}\\
 &&\Longrightarrow\hat{\xi}_{\mu}^{(1:2)}(\bm{n}_{\bm{q}},n_{\omega})=-\frac{\delta_{\mu1}}{\ell^d}\hat{u}(\bm{n}_{\bm{q}},n_{\omega})\frac{(1+\beta)(\tau_{\ell}\kappa^2q^4+\ui\tau_{\ell}\kappa\omega q^2)-2\ui\tau_{\ell}(1-\beta)\bm{g}\cdot\bm{q}}{2(\ui\omega+\kappa q^2)^2}\;.
\end{eqnarray*}
Upon antitransforming to the physical space and plugging into (\ref{kappa:0}) and (\ref{kappa:2}),
we get:
\begin{eqnarray} \label{kapparal}
 K_{\mu\nu}^{(:0)}&=&\displaystyle\kappa\delta_{\mu\nu}-\kappa\ell^d\!\mear\,\frac{\delta_{\mu1}\delta_{\nu1}}{\ell^{2d}}\times\nonumber\\
 &&\displaystyle\times\!\sum_{\bm{n}_{\bm{p}},\bm{n}_{\bm{q}}\in\mathbb{Z}^d}\sum_{n_{\Omega},n_{\omega}\in\mathbb{Z}}\!\hat{u}(\bm{n}_{\bm{p}},n_{\Omega})\hat{u}(\bm{n}_{\bm{q}},n_{\omega})\frac{\bm{p}\cdot\bm{q}}{(\ui\Omega+\kappa p^2)(\ui\omega+\kappa q^2)}\ue^{\ui[(\bm{p}+\bm{q})\cdot\bm{x}+(\Omega+\omega)t]}\nonumber\\
 &=&\displaystyle\kappa\left[\delta_{\mu\nu}+\delta_{\mu1}\delta_{\nu1}\!\sum_{\bm{n}_{\bm{q}}\in\mathbb{Z}^d}\sum_{n_{\omega}\in\mathbb{Z}}\!|\hat{u}(\bm{n}_{\bm{q}},n_{\omega})|^2\frac{q^2}{\omega^2+\kappa^2q^4}\right]\;,
\end{eqnarray}
\begin{eqnarray} \label{kp}
 K_{\mu\nu}^{(:2)}&=&\displaystyle\frac{\delta_{\mu1}\delta_{\nu1}}{\ell^d}\!\mear\,\bigg\{\tau_{\ell}(1-\beta)u^2(\bm{x},t)+\!\sum_{\bm{n}_{\bm{p}},\bm{n}_{\bm{q}}\in\mathbb{Z}^d}\sum_{n_{\Omega},n_{\omega}\in\mathbb{Z}}\!\hat{u}(\bm{p},\Omega)\hat{u}(\bm{q},\omega)\times\nonumber\\
 &&\left.\displaystyle\times\frac{(1+\beta)(\tau_{\ell}\kappa^2q^4+\ui\tau_{\ell}\kappa\omega q^2)-2\ui\tau_{\ell}(1-\beta)\bm{g}\cdot\bm{q}}{2(\ui\omega+\kappa q^2)^2}\ue^{\ui[(\bm{p}+\bm{q})\cdot\bm{x}+(\Omega+\omega)t]}\right\}\nonumber\\
 &=&\displaystyle\delta_{\mu1}\delta_{\nu1}\left\{\ell^{-d}\!\mear\,\tau_{\ell}(1-\beta)u^2(\bm{x},t)\right.\nonumber\\
 &&\displaystyle+\!\sum_{\bm{n}_{\bm{q}}\in\mathbb{Z}^d}\sum_{n_{\omega}\in\mathbb{Z}}\!|\hat{u}(\bm{n}_{\bm{q}},n_{\omega})|^2\frac{(1+\beta)(\tau_{\ell}\kappa^2q^4+\ui\tau_{\ell}\kappa\omega q^2)-2\ui\tau_{\ell}(1-\beta)\bm{g}\cdot\bm{q}}{2(\ui\omega+\kappa q^2)^2}\Bigg\}\;.
\end{eqnarray}
For generic parallel flows, gravity thus does appear in the $O(\mathrm{St})$ correction to the eddy diffusivity, (\ref{kp}),
and can in principle bring about \emph{either an enhancement or a depletion} of the tensorial component along the flow direction.
The tracer limit of the eddy diffusivity, (\ref{kapparal}), obviously coincides with the results of \citet{BCVV95},
and always shows a \emph{positive} correction \citep{PS05}
with respect to the Brownian isotropic counterpart in the absence of any flow.

\subsubsection{Steady Kolmogorov flow}
These results can be substantiated even more deeply by focusing on the steady Kolmogorov flow \citep{O83}
\begin{equation} \label{sK}
 u_{\mu}(\bm{x},t)=\delta_{\mu1}U\cos(q_{\ell}x_d)\;,
\end{equation}
with $q_{\ell}:=2\pi/\ell$, such that $\hat{u}(\bm{n}_{\bm{q}},n_{\omega})=U\delta_{n_{\omega},0}\delta_{n_{q_1},0}\ldots\delta_{n_{q_{d-1}},0}(\delta_{n_{q_d},1}+\delta_{n_{q_d},-1})/2$. Thus,
\begin{equation} \label{tensoriale}
 K_{\mu\nu}=\kappa\delta_{\mu\nu}+\frac{U^2}{2q_{\ell}^2\kappa}\delta_{\mu1}\delta_{\nu1}+\mathrm{St}\frac{U\ell(3-\beta)}{4}\delta_{\mu1}\delta_{\nu1}+O(\mathrm{St}^2)\;.
\end{equation}
In other words,
\begin{equation} \label{kk}
 K_{\parallel}:=K_{11}=\kappa\left[\left(1+\frac{\mathrm{Pe}^2}{8\pi^2}\right)+\frac{3-\beta}{4}\mathrm{Pe}\,\mathrm{St}+O(\mathrm{St}^2)\right]\;,
\end{equation}
i.e.\ the component of the eddy diffusivity parallel to the flow is \emph{enhanced} by the presence of the latter,
and gravity can play a role only in subleading corrections at higher orders in $\mathrm{St}$, independently of the cell orientation.
This enhancement is maximum at larger and larger $\mathrm{Pe}$ already at the tracer level,
and for very heavy particles taking into account the role of inertia at its leading order as well.
The \emph{further enhancement} due to inertia can easily be explained by reminding that, according to the Taylor formula,
the eddy diffusivity is loosely proportional to the particle Lagrangian autocorrelation time,
which is definitely longer for inertial particles than for tracers because of ballistic ``coherence'' or ``memory''
(notice that this mechanism provides an explanation for the present steady flow, but in general need not hold in unsteady cases).
On the other hand, no enhancement is present for a vanishing P\'eclet number, nor for bubbles.\\
The perpendicular component is \emph{not modified} within accuracy,
\begin{equation} \label{kperp}
 K_{\perp}:=\frac{K_{\mu\mu}-K_{\parallel}}{d-1}=\kappa\left[1+O(\mathrm{St}^2)\right]\;,
\end{equation}
and it is worth noticing that the $x_d$ axis does not introduce any preferential direction
in the $(d-1)$-dimensional subspace orthogonal to the flow, in which the eddy diffusivity is isotropic.

\subsubsection{Random-in-time Kolmogorov flow}
\begin{figure}
 \centering
 \includegraphics[height=8cm]{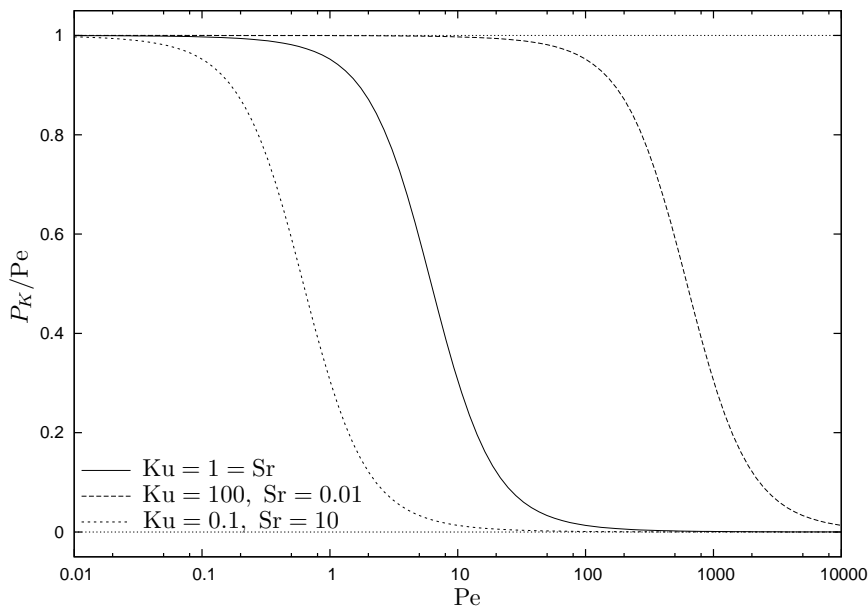}
 \caption{Ratio $P_K/\mathrm{Pe}$ as a function of $\mathrm{Pe}$ for different values of $\mathrm{Ku}$ and $\mathrm{Sr}$.
  Notice that the ratio is closer and closer to $1$ for large $\mathrm{Ku}$ and small $\mathrm{Sr}$, in which case (\ref{kpr}) tends to (\ref{kk}).}
 \label{fig:1}
\end{figure}

\begin{figure}
 \centering
 \includegraphics[height=4.5cm]{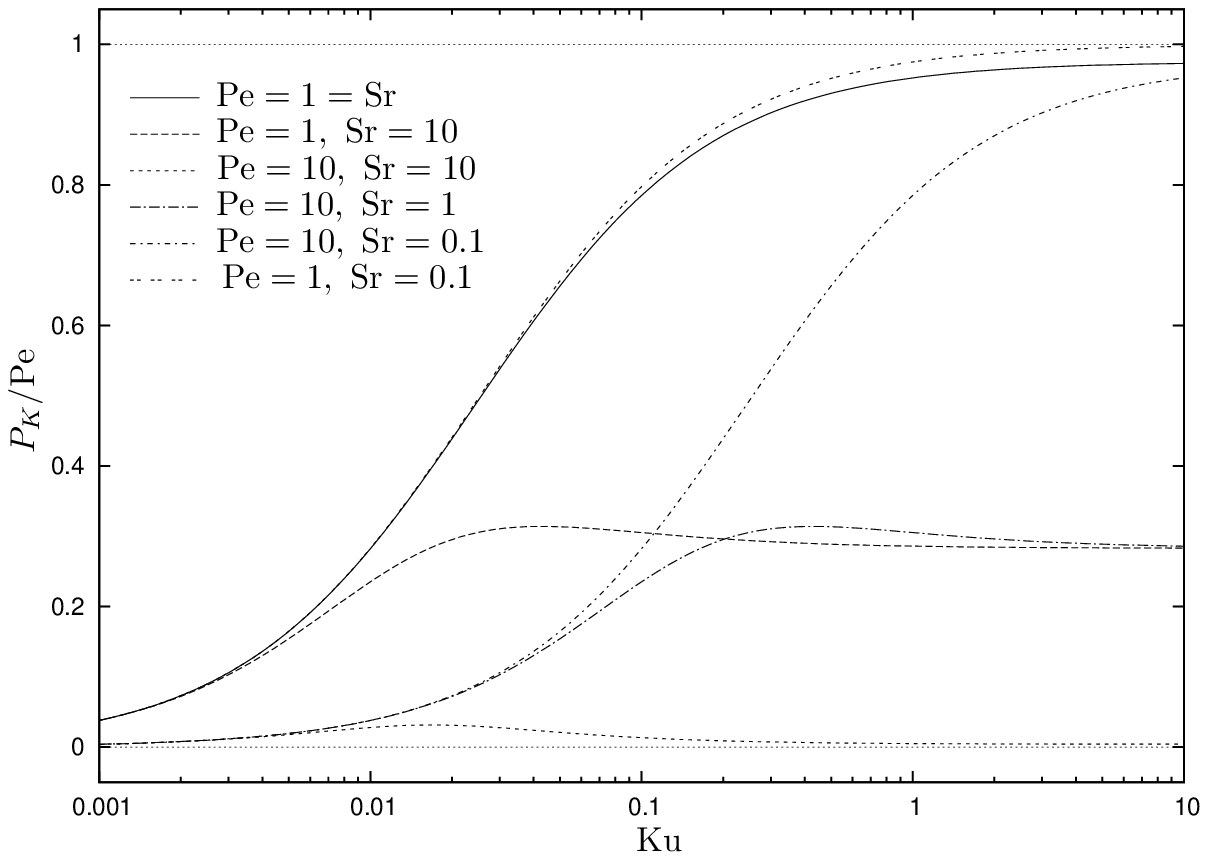}\hfill\includegraphics[height=4.5cm]{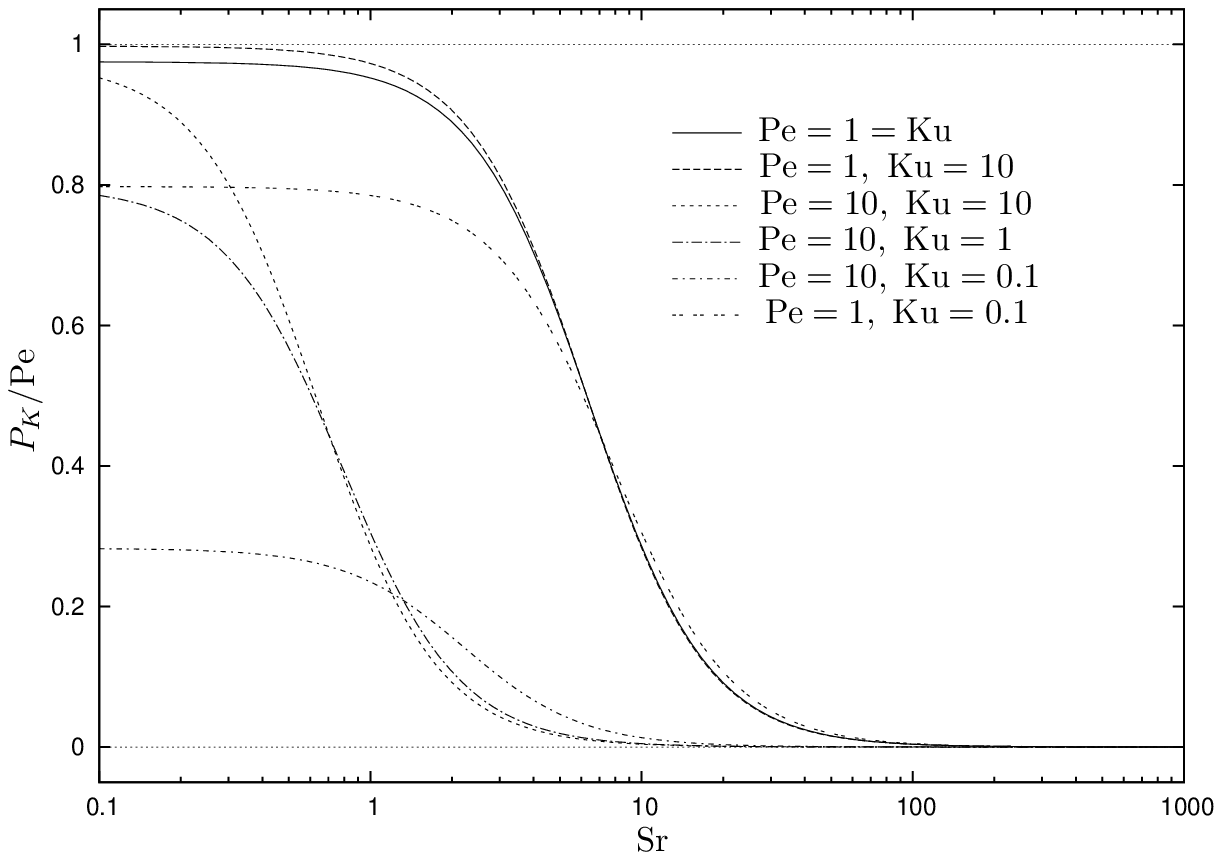}
 \caption{Left: ratio $P_K/\mathrm{Pe}$ as a function of $\mathrm{Ku}$ for different values of $\mathrm{Pe}$ and $\mathrm{Sr}$.
  Right: ratio $P_K/\mathrm{Pe}$ as a function of $\mathrm{Sr}$ for different values of $\mathrm{Pe}$ and $\mathrm{Ku}$.}
 \label{fig:2}
\end{figure}

As a further example, let us consider a random-in-time Kolmogorov flow,
\begin{equation} \label{ritK}
 u_{\mu}(\bm{x},t)=\delta_{\mu1}U\cos(q_{\ell}x_d)F(t)\;,
\end{equation}
such that the random temporal part is statistically stationary with correlation function \citep{MY75}
\begin{equation} \label{correlazione}
 C(t_{\Delta}):=\langle F(t+t_{\Delta})F(t)\rangle=\ue^{-\alpha|t_{\Delta}|}\cos(\gamma t_{\Delta})\;.
\end{equation}
This expression requires the introduction of two additional adimensional numbers,
\[\textrm{Kubo: }\mathrm{Ku}:=\frac{1}{\alpha\tau_{\ell}}\;,\qquad\textrm{Strouhal: }\mathrm{Sr}:=\frac{\gamma\tau_{\ell}}{2\pi}\;,\]
which give a measure of the vortex life time and of the recirculation degree, respectively, in advective-time units
(notice that, in the spirit of a comparison with the time-periodic case, $\gamma$ can roughly be identified with $2\pi/\mathcal{T}$).
The formalism we applied up to now still holds, with the proviso of dealing with a transform rather than a series in the Fourier frequency domain,
and of replacing any time average over a period with a statistical average $\langle\cdot\rangle$. We have indeed
\[\int_0^{\mathcal{T}}\!\frac{\ud t}{\mathcal{T}}\sum_{n_{\Omega},n_{\omega}\in\mathbb{Z}}\!\hat{u}(\bm{n}_{\bm{p}},n_{\Omega})\hat{u}(\bm{n}_{\bm{q}},n_{\omega})\ue^{\ui(\Omega+\omega)t}f(\omega,\Omega)\mapsto\frac{1}{\sqrt{2\pi}}\!\int_{\mathbb{R}}\!\ud\omega\,|\hat{u}_*(\bm{n}_{\bm{q}},\omega)|^2f(\omega,-\omega)\;,\]
with $|\hat{u}_*(\bm{n}_{\bm{q}},\omega)|^2=U^2\hat{F}(\omega)\delta_{n_{q_1},0}\ldots\delta_{n_{q_{d-1}},0}(\delta_{n_{q_d},1}+\delta_{n_{q_d},-1})/4$ and
\[\hat{F}(\omega)=\frac{\alpha}{\sqrt{2\pi}}\left[\frac{1}{\alpha^2+(\omega+\gamma)^2}+\frac{1}{\alpha^2+(\omega-\gamma)^2}\right]\;.\]
From (\ref{kapparal})--(\ref{kp}), the upshot is a modification only of the parallel eddy diffusivity:
\begin{equation} \label{kpr}
 K_{\parallel}=\kappa\left\{\left(1+\frac{\mathrm{Pe}^2}{8\pi^2}\frac{P_K}{\mathrm{Pe}}\right)+\left[\frac{1-\beta}{2}+\frac{1+\beta}{4}\frac{P_K}{\mathrm{Pe}}\right]\mathrm{Pe}\,\mathrm{St}+O(\mathrm{St}^2)\right\}\;.
\end{equation}
Here,
\begin{equation} \label{pks}
 P_K:=\frac{4\pi^2}{\tau_{\ell}}\frac{\alpha+q_{\ell}^2\kappa}{(\alpha+q_{\ell}^2\kappa)^2+\gamma^2}=\frac{\mathrm{Pe}^{-1}+\mathrm{Ku}^{-1}/4\pi^2}{(\mathrm{Pe}^{-1}+\mathrm{Ku}^{-1}/4\pi^2)^2+\mathrm{Sr}^2/4\pi^2}
\end{equation}
can be identified as a ``modified'' P\'eclet number, which gets back its ``original'' value $\mathrm{Pe}$ in the steady case $\alpha\to0\gets\gamma$,
i.e.\ for infinite $\mathrm{Ku}$ and vanishing $\mathrm{Sr}$ (in which case (\ref{kpr}) coincides with (\ref{kk})).
In general, a non-vanishing decaying rate $\alpha$ (finite $\mathrm{Ku}$) helps increasing the role of the molecular diffusivity
in the expression of the eddy diffusivity, while a finite recirculation degree $\gamma$ (non-vanishing $\mathrm{Sr}$)
contributes to reducing the parametre $P_K$.\\
The outcome is that, at the tracer level ($\mathrm{St}\to0$ in (\ref{kpr})), the correction with respect to the Brownian value
is always \emph{positive}, but its magnitude is smaller with respect to the steady-Kolmogorov correction (\ref{kk})
because of a multiplicative factor $P_K/\mathrm{Pe}$ ($\le1$ and investigated in figures \ref{fig:1}--\ref{fig:2}).\\
For what concerns the role of inertia at its leading order,
again the magnitude of the correction is smaller with respect to the steady-Kolmogorov case (note that $(3-\beta)/4=(1-\beta)/2+(1+\beta)/4$,
for a comparison with (\ref{kk})), but its sign turns out to show a diffusivity \emph{depletion for} particles light enough,
as the coefficient in square brackets in (\ref{kpr}) is negative for $\beta>\beta^*$. Here, the critical value
\[\beta^*:=\frac{3}{2-P_K/\mathrm{Pe}}\in[1.5,3]\]
is a growing function of $P_K$, and thus gets back its ``steady'' value $3$ for infinite $\mathrm{Ku}$ and vanishing $\mathrm{Sr}$.\\

\begin{figure}
 \centering
 \includegraphics[height=8cm]{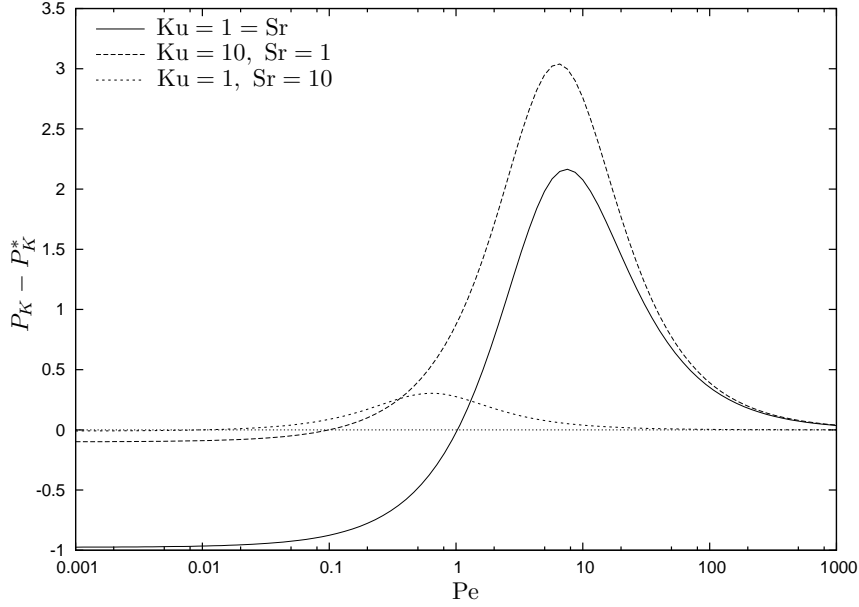}
 \caption{Sign of the interference (\ref{inter}) as a function of $\mathrm{Pe}$ for different values of $\mathrm{Ku}$ and $\mathrm{Sr}$.
  Note that the two dashed lines correspond to the same value of the product $\mathrm{Ku}\,\mathrm{Sr}$ but give different results, which means
  that the ratio $\gamma/\alpha$ from (\ref{correlazione}) alone in this case is not sufficient to determine the nature of the interference.}
 \label{fig:3}
\end{figure}

An interesting quantity to investigate is the \emph{interference} \citep{MV97,CCMVV98}
between the different physical mechanisms into play, which possess different intrinsic time scales
(advective, diffusive, inertial, $1/\alpha$, $1/\gamma$) and can thus interfere constructively
--- to give a resonance --- or destructively. We then compute the quantity
\begin{equation} \label{inter}
 \mathcal{K}:=K_{\parallel}-\kappa-\lim_{\mathrm{Pe}\to\infty}K_{\parallel}=U\ell\left[\frac{1}{8\pi^2}(P_K-P_K^*)+\frac{1+\beta}{4}\frac{P_K}{\mathrm{Pe}}\mathrm{St}+O(\mathrm{St}^2)\right]\;,
\end{equation}
where
\begin{equation} \label{picappa}
 P_K^*:=\lim_{\mathrm{Pe}\to\infty}P_K=\frac{\mathrm{Ku}^{-1}}{\mathrm{Ku}^{-2}/4\pi^2+\mathrm{Sr}^2}\;.
\end{equation}
A positive or negative sign of $\mathcal{K}$ then indicates a constructive or destructive interference
of the Brownian diffusivity with the other mechanisms.\\
At the tracer level, this corresponds to investigate the sign of $(P_K-P_K^*)$, which is done in figure \ref{fig:3}
and shows the possible presence of both alternatives. Analytically, one finds a critical value of the P\'eclet number,
\[\mathrm{Pe}^*:=\frac{\mathrm{Ku}^{-1}}{\mathrm{Sr}^2-\mathrm{Ku}^{-2}/4\pi^2}\;,\]
such that the interference is constructive for $\mathrm{Pe}>\mathrm{Pe}^*$ and destructive otherwise
(the interference is always destructive if $\mathrm{Pe}^*$ turns out to be negative, e.g.\ for vanishing $\mathrm{Sr}$).\\
Then, one can also take inertia into account at its leading order, which always carries a positive addend in $\mathcal{K}$,
whose magnitude increases linearly with both $\beta$ and $\mathrm{St}$. Starting from situations where $\mathcal{K}<0$
in the tracer approximation $\mathrm{St}\to0$, it is thus interesting to investigate for what ranges of $\beta$ and $\mathrm{St}$
the interference becomes positive: an example is sketched in table \ref{tab:1}.
\begin{table}
 \begin{center}
  \begin{tabular}{cccc}
   $\mathrm{St}<3.05\;10^{-4}$ & $\mathrm{St}=10^{-3.5}$ & $\mathrm{St}=10^{-3}$ & $\mathrm{St}>1.22\;10^{-3}$\\
   \hline
   $\nexists\beta$ & $\beta>2.86$ & $\beta>0.22$ & $\forall\beta$\\
   (no type) & (only bubbles) & (also tracers) & (every particle)
  \end{tabular}
 \end{center}
 \caption{Focusing on the case $\mathrm{Pe}=\mathrm{Ku}=\mathrm{Sr}=1$, i.e.\ a situation where the interference is destructive
  in the tracer approximation $\mathrm{St}\to0$, the table reports (for different, small values of $\mathrm{St}$)
  the ranges of $\beta$ which turn the interference constructive upon taking inertia into account at its leading order.}
 \label{tab:1}
\end{table}

\subsection{Expansion at small inertia and finite terminal velocity}

Let us now investigate the case of finite terminal velocity (\ref{Maxey:limit})
in the framework of parallel flows,
i.e.\ let us consider (\ref{csitilde}) instead of (\ref{csi0BIS}).
We then obtain a modified version of (\ref{csi0fourier}):
\begin{eqnarray*}
 \lefteqn{\left[\frac{\partial}{\partial t}+(1-\beta)S_{F}U\bm{G}\cdot\frac{\partial}{\partial\bm{x}}-\kappa\Delta_{\bm{x}}\right]\tilde{\xi}_{\mu}^{(1:0)}=-\frac{\delta_{\mu1}}{\ell^d}u}\\
 &&\Longrightarrow\hat{\tilde{\xi}}_{\mu}^{(1:0)}(\bm{n}_{\bm{q}},n_{\omega})=-\frac{\delta_{\mu1}}{\ell^d}\frac{\hat{u}(\bm{n}_{\bm{q}},n_{\omega})}{\ui\omega+\ui(1-\beta)S_{F}U\bm{G}\cdot\bm{q}+\kappa q^2}\;.
\end{eqnarray*}
From (\ref{kappatilde}),
we deduce that (\ref{kapparal}) is to be modified into:
\begin{eqnarray*}
 K_{\mu\nu}^{(:0)}\mapsto\tilde{K}_{\mu\nu}^{(:0)}&=&\displaystyle\kappa\delta_{\mu\nu}-\kappa\ell^d\!\mear\frac{\delta_{\mu1}\delta_{\nu1}}{\ell^{2d}}\!\sum_{\bm{n}_{\bm{p}},\bm{n}_{\bm{q}}\in\mathbb{Z}^d}\sum_{n_{\Omega},n_{\omega}\in\mathbb{Z}}\!\hat{u}(\bm{n}_{\bm{p}},n_{\Omega})\hat{u}(\bm{n}_{\bm{q}},n_{\omega})\times\\
 &&\displaystyle\times\frac{(\bm{p}\cdot\bm{q})\ue^{\ui[(\bm{p}+\bm{q})\cdot\bm{x}+(\Omega+\omega)t]}}{[\ui\Omega+\ui(1-\beta)S_{F}U\bm{G}\cdot\bm{p}+\kappa p^2][\ui\omega+\ui(1-\beta)S_{F}U\bm{G}\cdot\bm{q}+\kappa q^2]}\\
 &=&\displaystyle\kappa\left[\delta_{\mu\nu}+\delta_{\mu1}\delta_{\nu1}\!\sum_{\bm{n}_{\bm{q}}\in\mathbb{Z}^d}\sum_{n_{\omega}\in\mathbb{Z}}\!\frac{|\hat{u}(\bm{n}_{\bm{q}},n_{\omega})|^2q^2}{[\omega+(1-\beta)S_{F}U\bm{G}\cdot\bm{q}]^2+\kappa^2q^4}\right]\;.
\end{eqnarray*}

\subsubsection{Steady Kolmogorov flow}
Let us now impose once again the steady Kolmogorov flow (\ref{sK}) and, instead of (\ref{tensoriale}), we get
\[K_{\mu\nu}\mapsto\tilde{K}_{\mu\nu}=\kappa\left[\delta_{\mu\nu}+\delta_{\mu1}\delta_{\nu1}\frac{U^2}{2U^2(1-\beta)^2S_F^2\cos^2\theta+2q_{\ell}^2\kappa^2}\right]+O(\mathrm{St})\;,\]
where $\theta$ is the angle between gravity and the $x_d$ direction. In other words, (\ref{kk})--(\ref{kperp}) now become
\begin{equation} \label{kkm}
 K_{\parallel}\mapsto\tilde{K}_{\parallel}=\kappa\left[1+\frac{\mathrm{Pe}^2}{8\pi^2+2\mathrm{Pe}^2(1-\beta)^2S_F^2\cos^2\theta}+O(\mathrm{St})\right]\;,
\end{equation}
\begin{equation} \label{kpe}
 K_{\perp}\mapsto\tilde{K}_{\perp}=\kappa\left[1+O(\mathrm{St})\right]\;,
\end{equation}
Comparing (\ref{kkm}) with the corresponding result for vanishing terminal velocity, (\ref{kk}), we deduce that the effect of gravity at
$O(\mathrm{St}^0)$ is to \emph{reduce the enhancement} of the parallel eddy diffusivity caused by the presence of the flow with respect
to the Brownian value. This reduction of the eddy-diffusivity enhancement is most pronounced when the $x_d$ direction (on which the Kolmogorov flow
depends sinusoidally) is vertically aligned --- independently whether upwards or downwards ---, which also means that no enhancement reduction
is present if the flow is vertical ($x_1\parallel\bm{G}\Rightarrow x_d\perp\bm{G}\Rightarrow\cos\theta=0$). For what concerns the eddy diffusivity
perpendicular to the flow, a comparison between (\ref{kperp}) and (\ref{kpe}) shows \emph{no change} at $O(\mathrm{St}^0)$,
but a finite terminal velocity can in general move the effect of gravity from the original $O(\mathrm{St}^2)$ down to $O(\mathrm{St}^1)$.\\
Alternatively, one may study the eddy diffusivity in the vertical and any horizontal direction;
their expressions coincide with (\ref{kkm}) but with the fraction in square brackets multiplied by $\cos^2\Theta$
and $(d-1)^{-1}\sin^2\Theta$ respectively, where $\Theta$ is the angle between the vertical and the flow directions.

\subsubsection{Random-in-time Kolmogorov flow}
\begin{figure}
 \centering
 \includegraphics[height=4.5cm]{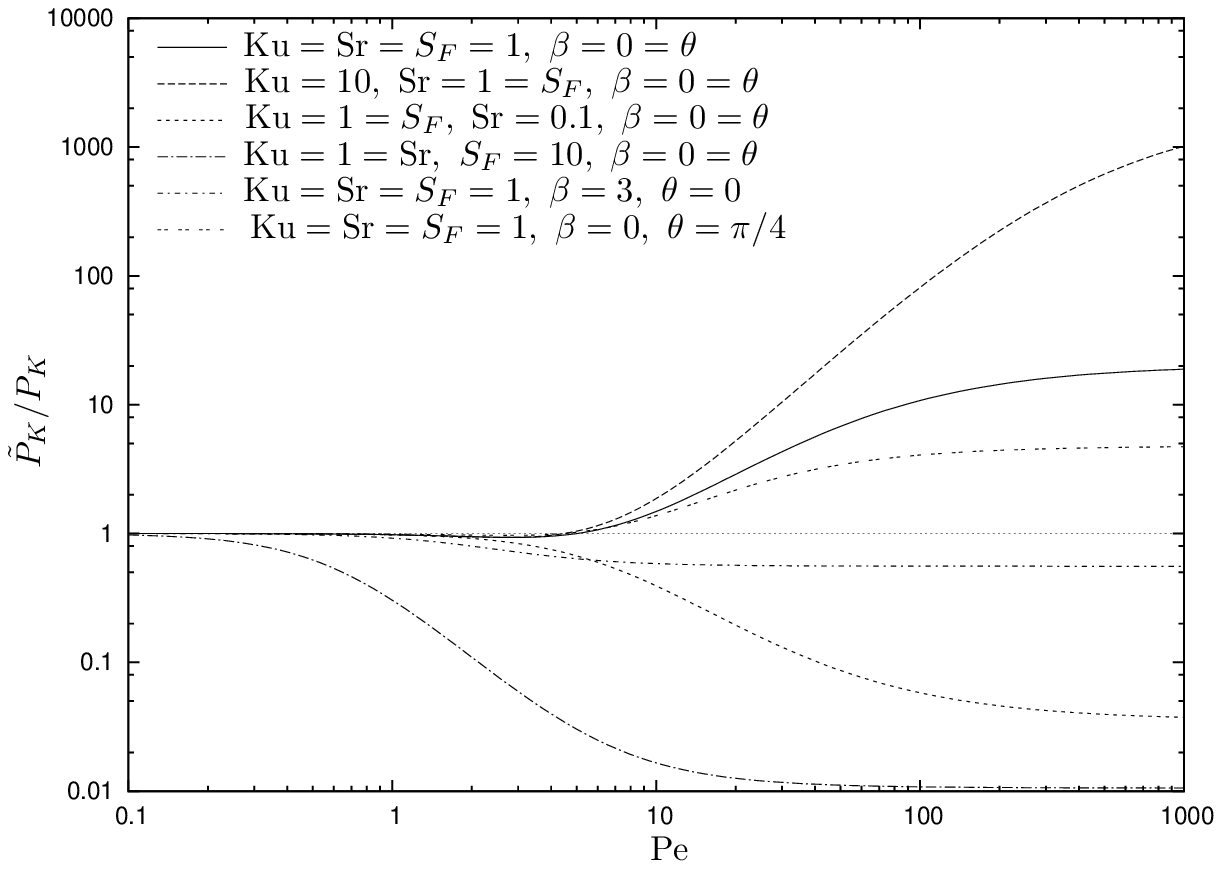}\hfill\includegraphics[height=4.5cm]{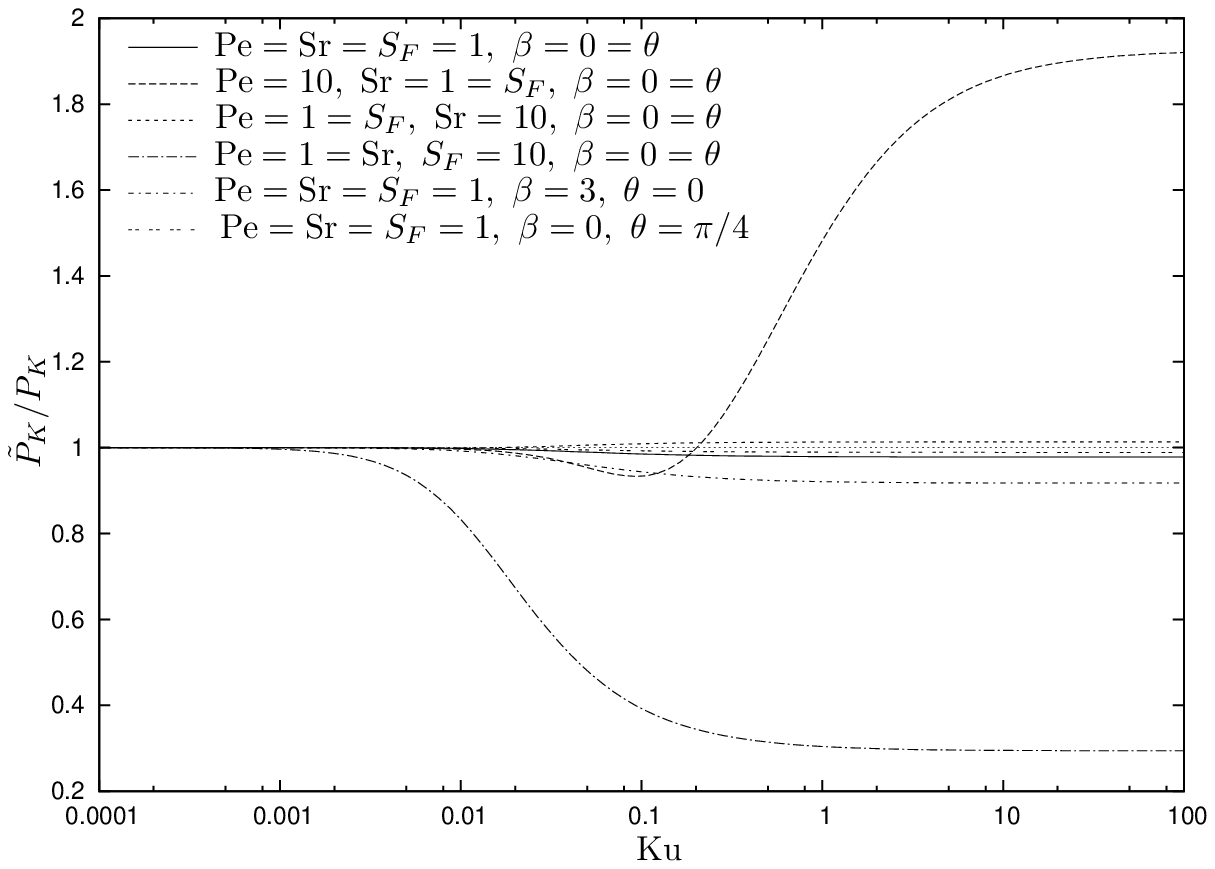}
 \caption{Left: ratio $\tilde{P}_K/P_K$ as a function of $\mathrm{Pe}$ for different values of $\mathrm{Ku}$, $\mathrm{Sr}$, $S_F$, $\beta$ and
  $\theta$; the ratio is closer and closer to $1$ for small $S_F$, $(1-\beta)$ or $(\pi/2-\theta)$, in which case (\ref{kprk}) tends to (\ref{kpr}).
  Right: ratio $\tilde{P}_K/P_K$ as a function of $\mathrm{Ku}$ for different values of $\mathrm{Pe}$, $\mathrm{Sr}$, $S_F$, $\beta$ and $\theta$.}
 \label{fig:4}
\end{figure}

\begin{figure}
 \centering
 \includegraphics[height=4.5cm]{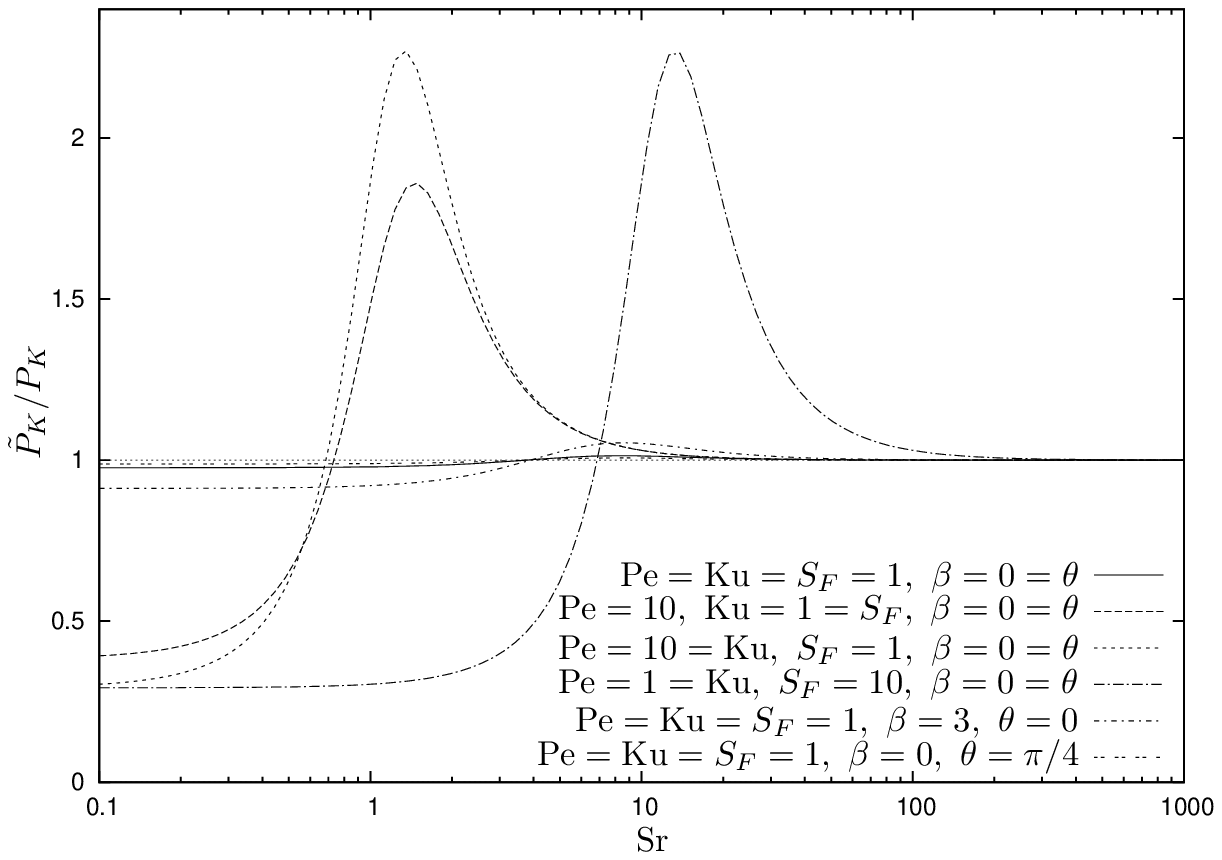}\hfill\includegraphics[height=4.5cm]{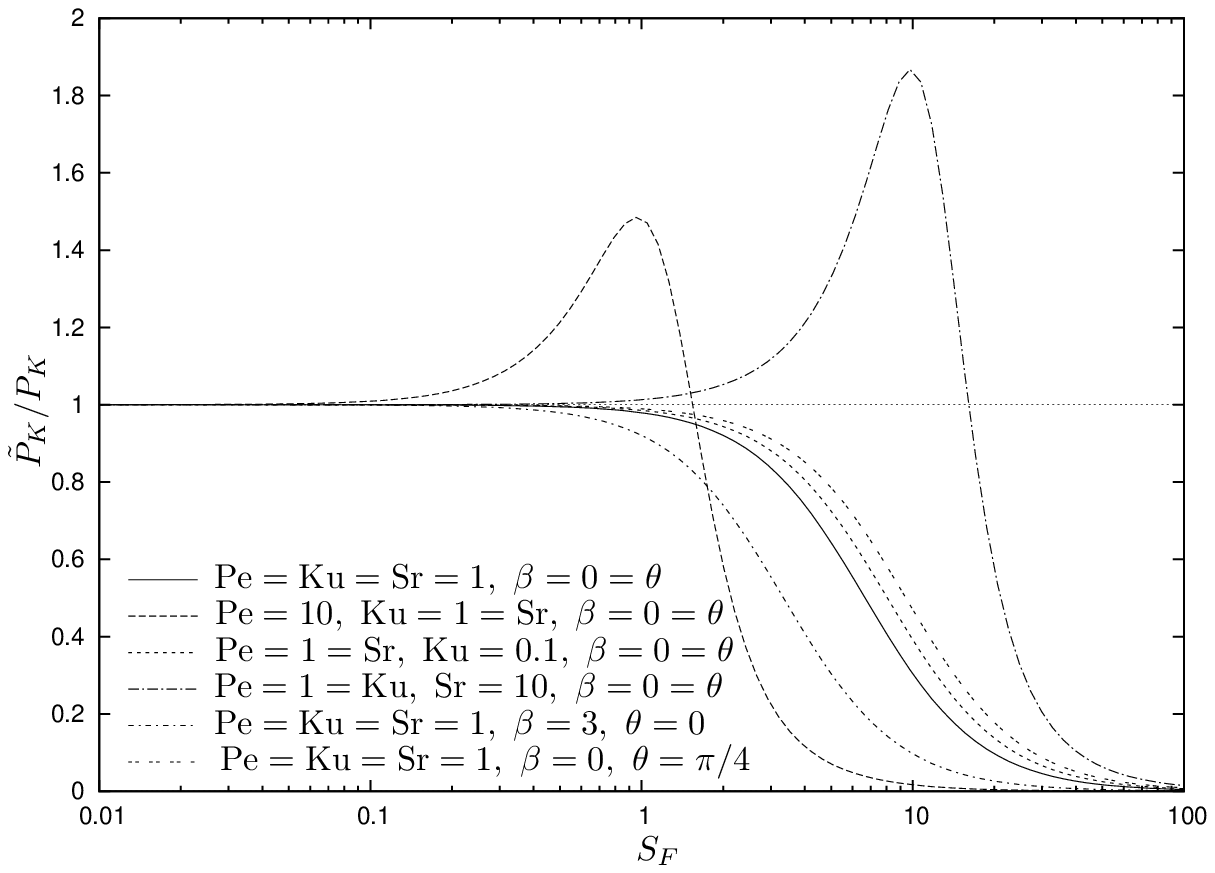}
 \caption{Left: ratio $\tilde{P}_K/P_K$ as a function of $\mathrm{Sr}$ for different values of $\mathrm{Pe}$, $\mathrm{Ku}$, $S_F$, $\beta$ and
  $\theta$. Right: ratio $\tilde{P}_K/P_K$ as a function of $S_F$ for different values of $\mathrm{Pe}$, $\mathrm{Ku}$, $\mathrm{Sr}$, $\beta$ and
  $\theta$.}
 \label{fig:5}
\end{figure}

\begin{figure}
 \centering
 \includegraphics[height=4.5cm]{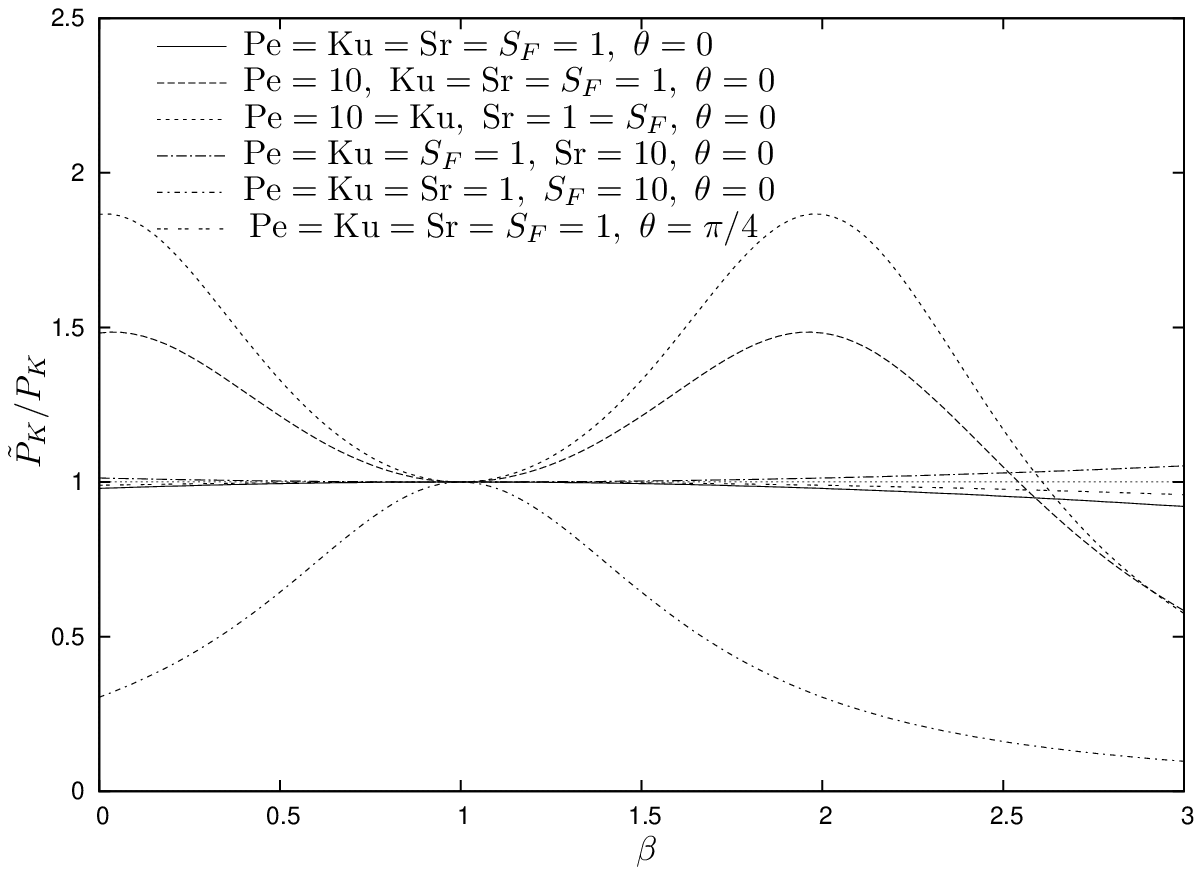}\hfill\includegraphics[height=4.5cm]{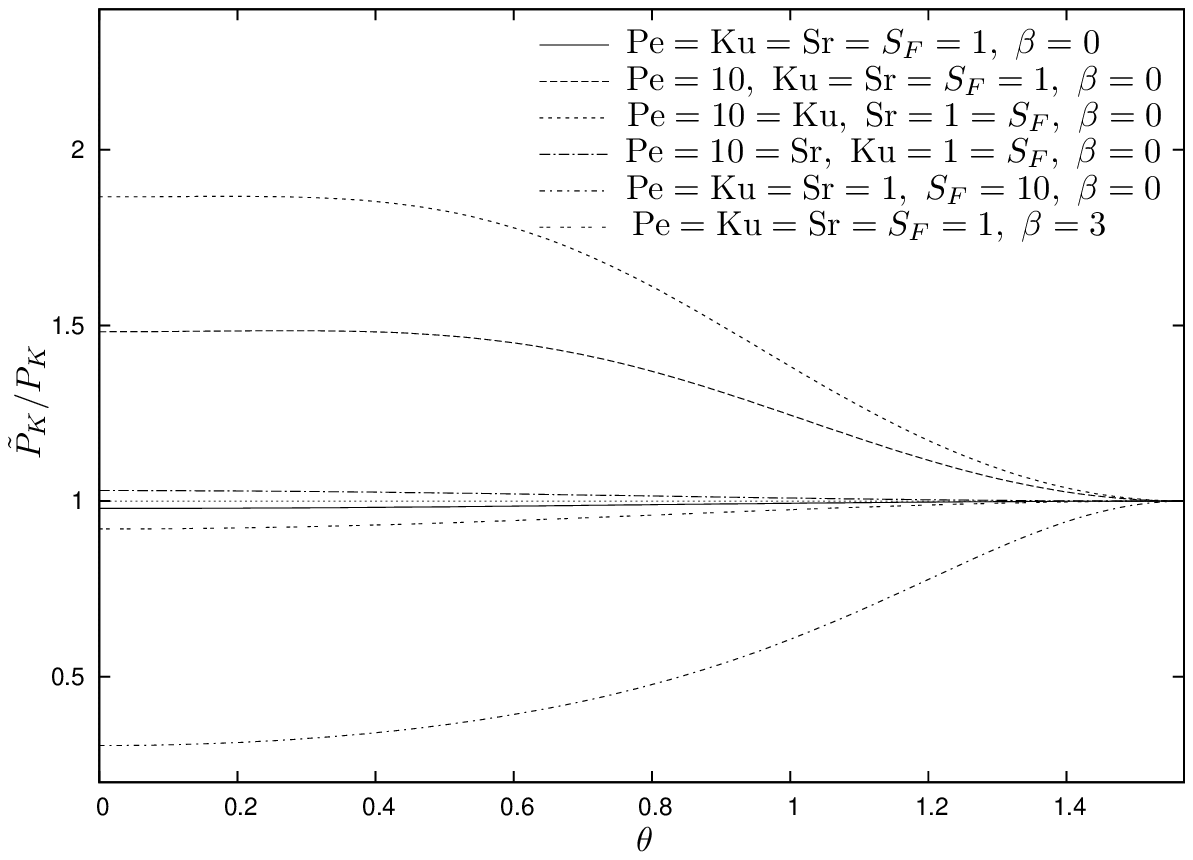}
 \caption{Left: ratio $\tilde{P}_K/P_K$ as a function of $\beta$ for different values of $\mathrm{Pe}$, $\mathrm{Ku}$, $\mathrm{Sr}$, $S_F$ and
  $\theta$. Right: ratio $\tilde{P}_K/P_K$ as a function of $\theta$ for different values of $\mathrm{Pe}$, $\mathrm{Ku}$, $\mathrm{Sr}$, $S_F$ and
  $\beta$.}
 \label{fig:6}
\end{figure}

If, on the contrary, the random Kolmogorov flow (\ref{ritK})--(\ref{correlazione}) is imposed,
the eddy diffusivity is again modified only in its parallel component, according not to (\ref{kpr}) but rather to
\begin{equation} \label{kprk}
 K_{\parallel}\mapsto\tilde{K}_{\parallel}=\kappa\left[1+\frac{\mathrm{Pe}^2}{8\pi^2}\frac{\tilde{P}_K}{\mathrm{Pe}}+O(\mathrm{St})\right]\;.
\end{equation}
Here, (\ref{pks}) becomes
\[P_K\mapsto\tilde{P}_K:=\frac{4\pi^2}{\tau_{\ell}}\left[\frac{1}{(\alpha+q_{\ell}^2\kappa)^2+{\Gamma_+}^2}+\frac{1}{(\alpha+q_{\ell}^2\kappa)^2+{\Gamma_-}^2}\right]\frac{\alpha+q_{\ell}^2\kappa}{2}\;,\]
with
\[\Gamma_{\pm}:=\gamma\pm Uq_{\ell}(1-\beta)S_F\cos\theta\;.\]
This latter expression shows that sedimentation
contributes to change the flow characteristic frequency $\gamma$ as viewed (or ``sampled'') by the particles,
but in a manner symmetric for up/down reflections ($\theta\mapsto\pi-\theta$) because of the symmetrization present in $\tilde{P}_K$
(clearly no modification occurs for neutral particles $\beta=1$, nor for the orthogonal case $\theta=\pi/2$).
In the steady case $\alpha=0=\gamma$, expression (\ref{kprk}) obviously coincides with (\ref{kkm}).\\
Another interesting comparison for (\ref{kprk}) is with the corresponding result (\ref{kpr}) for vanishing terminal velocity (limit
$S_F\to0$); then the ratio $\tilde{P}_K/P_K$ has to be investigated, which is done in figures \ref{fig:4}--\ref{fig:6}: the upshot is that, in the
framework of the random-in-time Kolmogorov flow, gravity can \emph{either increase or decrease the enhancement} of the parallel eddy diffusivity,
according to the values of the other parametres.\\

\begin{figure}
 \centering
 \includegraphics[height=8cm]{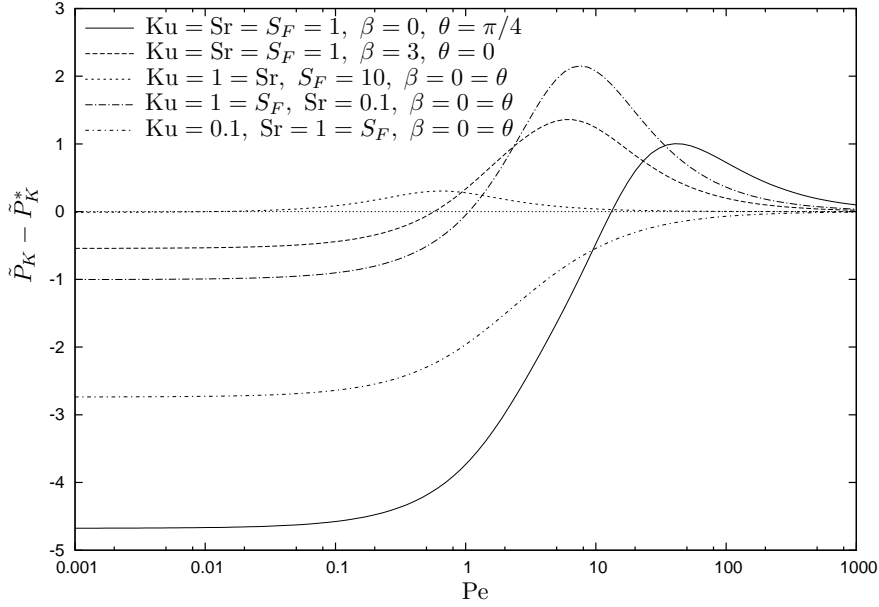}
 \caption{Sign of the interference (\ref{interK}) as a function of $\mathrm{Pe}$
  for different values of $\mathrm{Ku}$, $\mathrm{Sr}$, $S_F$, $\beta$ and $\theta$.}
 \label{fig:7}
\end{figure}

\begin{figure}
 \centering
 \includegraphics[height=8cm]{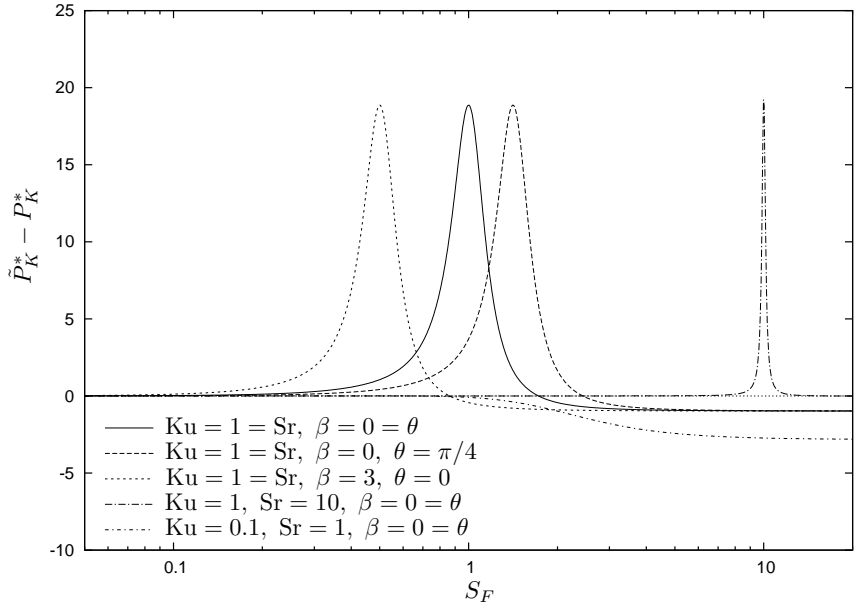}
 \caption{Sign of the interference (\ref{interG}) as a function of $S_F$ for different values of $\mathrm{Ku}$, $\mathrm{Sr}$, $\beta$ and $\theta$.}
 \label{fig:8}
\end{figure}

Let us now analyse the problem of interference again. From the usual point of view, this implies investigating not (\ref{inter}) but rather
\begin{equation} \label{interK}
 \mathcal{K}\mapsto\tilde{\mathcal{K}}:=\tilde{K}_{\parallel}-\kappa-\lim_{\mathrm{Pe}\to\infty}\tilde{K}_{\parallel}=U\ell\left[\frac{1}{8\pi^2}(\tilde{P}_K-\tilde{P}_K^*)+O(\mathrm{St})\right]\;,
\end{equation}
with (\ref{picappa}) now becoming
\begin{eqnarray*}
 &&P_K^*\mapsto\tilde{P}_K^*:=\lim_{\mathrm{Pe}\to\infty}\tilde{P}_K\\
 &&\quad=\frac{1}{2}\left\{\frac{\mathrm{Ku}^{-1}}{\mathrm{Ku}^{-2}/4\pi^2+[\mathrm{Sr}+(1-\beta)S_F\cos\theta]^2}+\frac{\mathrm{Ku}^{-1}}{\mathrm{Ku}^{-2}/4\pi^2+[\mathrm{Sr}-(1-\beta)S_F\cos\theta]^2}\right\}\;.
\end{eqnarray*}
A positive or negative sign of $\tilde{\mathcal{K}}$ then indicates a constructive or destructive interference of the \emph{Brownian
diffusivity} with the other mechanisms: this corresponds to investigating the sign of $(\tilde{P}_K-\tilde{P}_K^*)$ as a function of
$\mathrm{Pe}$ ($\propto\kappa^{-1}$), which is done in figure \ref{fig:7} and shows the possible presence of both alternatives.\\
However, one could also wonder how \emph{gravity} interferes with the other mechanisms. Therefore, one can study the sign of the quantity
\begin{equation} \label{interG}
 \mathcal{K}_*:=\left.\left(\tilde{K}_{\parallel}-\kappa-\tilde{K}_{\parallel}|_{g=0}\right)\right|_{\kappa=0}=\lim_{\mathrm{Pe}\to\infty}\left(\tilde{K}_{\parallel}-K_{\parallel}\right)=U\ell\left[\frac{1}{8\pi^2}(\tilde{P}_K^*-P_K^*)+O(\mathrm{St})\right]\;,
\end{equation}
i.e.\ of $(\tilde{P}_K^*-P_K^*)$ as a function of $S_F\propto||\bm{g}||$ (where the Brownian diffusivity must be neglected for sake of
consistency). This is done in figure \ref{fig:8} and again shows the possibility of having either destructive or constructive interference,
with the relevant annotation that the latter can now be very pronounced near well-defined resonance peaks at critical values
\[S_F^*:=\frac{\mathrm{Sr}}{|1-\beta|\cos\theta}\]
(except for small values of $\mathrm{Ku}$, which make the denominators in $\tilde{P}_K^*$ large anyway).

Notice that the situation analysed in the present subsection is a generalization of the one already attacked in \citet{M97,MV97},
with the role of the mean, large-scale drift played here by sedimentation.

\section{Conclusions and perspectives} \label{sec:conc}

We have investigated the asymptotic (in space and time) behaviour
of the concentration field of inertial particles of arbitrary mass
density carried by an incompressible flow field (static or time dependent)
under the action of gravity.

Exploiting the multiple-scale perturbative expansion
in the scale-separation parametre, we found a diffusive equation
for the asymptotic dynamics of the particle concentration field, involving
a positive definite eddy-diffusivity tensor. The method reduces the computation
of this tensor to the solution of an auxiliary differential problem
valid for any Stokes, Froude and P\'eclet numbers.
The general results have been greatly simplified focusing on two relevant
limits of vanishing inertia: one with vanishing sedimenting velocity,
the other maintaining a finite value of the latter quantity.\\
As also happening in the simpler problem of inertialess-particle transport,
the determination of the eddy-diffusivity tensor is in general possible
only exploiting numerical strategies. Here the problem is particularly
evident owing to the larger (compared to the inertialess case) phase space.
Nevertheless, we were able to select a relevant class of carrier flows
(parallel flows, both static and time-dependent)
where, after some manipulations, one arrives to exact, explicit expressions
for the eddy diffusivities. The latter expressions
allow us to extricate a highly
nontrivial role played by gravity and inertia
on the diffusive behaviour, where regimes of both
enhanced and reduced diffusion can be observed depending
on the detailed structure of the flow.

Here is a summary of our findings for the parallel Kolmogorov flows, highlighting
the nontrivial role of inertia and gravity on the diffusion process in the direction parallel to the flow.
\begin{itemize}
 \item An enhancement of the diffusivity is always found, and the P\'eclet number always plays a positive role in this respect (more precisely, analysing the ratio $K_{\parallel}/\kappa$). However, if one compares the situations in which other mechanisms --- such as flow time dependence or sedimentation or both --- are present, with the corresponding situations in which the latter are absent, one can find an enhancement reduction which is more or less pronounced at different $\mathrm{Pe}$. By introducing the concept of interference, one concludes that Brownian diffusivity interferes constructively/destructively with the other mechanisms for P\'eclet numbers larger/smaller than a critical value, respectively, but the latter can in principle be made arbitrarily small or large. These conclusions hold in general, irrespectively of considering a tracer or an inertial particle at small inertia, endowed with vanishing or finite sedimentation.
 \item The role of inertia (Stokes number) should be taken into account together with the coefficient $\beta$ based on the density ratio. The diffusivity enhancement indeed shows a leading correction proportional to $\mathrm{St}$, which is always positive (and decreasing with $\beta$) for a steady flow, but which can become negative for a random time-dependent flow if the particles are light enough. This enhancement reduction, while passing from the former kind of flow to the latter, is a decreasing function of the Strouhal number --- i.e.\ vortex frequency or recirculation degree --- and a growing function (loosely speaking, except for some peculiar cases or ranges) of the Kubo number --- i.e.\ the life time of the structures before decay. (Also notice that $\mathrm{Ku}$ always appears in the form of a harmonic sum with $\mathrm{Pe}$, as resistances in parallel.) $\mathrm{St}$ and $\beta$ also contribute linearly and positively to the above-mentioned interference.
 \item Gravity (Froude number) plays a significant role only through sedimentation, i.e.\ when the particle terminal velocity ($S_F$, modulo a prefactor) remains finite even for vanishing inertia. In that case, the key quantity is the combination $(1-\beta)S_F\cos\theta$, $\theta$ being the angle between the vertical and the $x_d$ direction upon which the flow depends sinusoidally (and NOT the $x_1$ direction along which the flow points: this distinction is crucial in the three-dimensional case). This key parametre brings about a monotonically growing reduction of the diffusivity enhancement in the steady case. In the random time-dependent case, this enhancement reduction: coincides with the vanishing-sedimentation value for a vanishing deviation of $\beta$ from $1$, or of $S_F$ from $0$, or of $\theta$ from $\pi/2$; is maximum for some intermediate deviation, corresponding to a resonance with the flow frequency $\mathrm{Sr}$; and vanishes for very large deviations, i.e.\ infinite $S_F$. Such a behaviour is summarised by the presence of strong resonance peaks denoting constructive interference of gravity with the other mechanisms; this interference becomes destructive for $S_F$ large enough.\\
\end{itemize}

We thank Antonio Celani for useful discussions and suggestions.
MMA acknowledges financial support from ``Institut de Radioprotection et de S\^uret\'e Nucl\'eaire'',
and PMG from ``Center of Excellence AnDy of the Finnish Academy''.

\appendix

\section{Details of the resolution of (\ref{prima})--(\ref{quarta}) for $\bm{\psi}$} \label{ap:det}

\subsection{Phase-space density $\psi_0\equiv p^{(0)}$} \label{sec:MA08p}

The Stokes-number expansion of the phase-space density $p^{(0)}$ was studied in details in
\citet{MA08}, the main results whereof we hereby shortly revisit by couching them
into our notation. The results the Ornstein--Uhlenbeck process recalled in \S~\ref{ap:OU}
imply that any integrable initial co-velocity distribution $\bar{\rho}_{v}$ evolving
in $t^{\ddag}$ according to (\ref{smin:OU}) converges exponentially to the
Maxwell--Boltzmann distribution
\begin{equation} \label{MA08p:MB}
 p^{(0:0)}(\bm{x},\bm{v},t)=\frac{\ue^{-v^2/2\sigma_{\ell}^2}}{(2\pi\sigma_{\ell}^2)^{d/2}}\xi^{(0:0)}(\bm{x},t)\equiv\chi^{0;\sigma_{\ell}}(\bm{v})\xi^{(0:0)}(\bm{x},t)\;.
\end{equation}
The function $\xi^{(0:0)}$ is a constant with respect to the fast variables
$(\bm{v},t^{\ddag})$. We leverage it to impose the solvability conditions of the
perturbative hierarchy starting at (\ref{MA08p:MB}).
Higher-order corrections are most conveniently evaluated observing that the
grading (\ref{ap:OU_grading}) and recursion (\ref{ap:OU_recursion}) relations,
enjoyed by the Hermite polynomials $H^{n;\sigma_{\ell}}_{\dots}$'s and their
$\mathbb{L}^2(\mathbb{R}^{d})$-orthogonal duals $\chi^{n;\sigma_{\ell}}_{\dots}$'s, imply that for any $f(\bm{x})$
\begin{eqnarray} \label{MA08p:L1recursion}
 \lefteqn{\mathcal{L}^{(0:1)}\chi^{n;\sigma_{\ell}}_{\alpha_1\dots\alpha_n}(\bm{v})f(\bm{x})=}\nonumber\\
 &&-\chi^{n+1;\sigma_{\ell}}_{\alpha_1\dots\alpha_n\alpha_{n+1}}(\bm{v})\mathcal{M}_{\alpha_{n+1}}f(\bm{x})+\sigma_{\ell}\sqrt{\frac{2}{\mathrm{Pe}}}\,\sum_{k=1}^n\chi^{n-1;\sigma_{\ell}}_{\alpha_1\dots\alpha_{k-1}\alpha_{k+1}\dots\alpha_n}(\bm{v})\frac{\partial}{\partial x_{\alpha_k}}f(\bm{x})
\end{eqnarray}
and%
\footnote{throughout the article, unless explicitly stated,
 the Einstein convention of implicit summation holds over repeated indices
 $\alpha_i=1,\dots,d$ ($i=1,\dots,n$, for any $n$).}
\begin{equation} \label{MA08p:L3recursion}
 \mathcal{L}^{(0:3)}\chi^{n;\sigma_{\ell}}_{\alpha_1\dots\alpha_n}(\bm{v})f(\bm{x})=-\sqrt{\frac{\mathrm{Pe}}{2}}\,\frac{(1-\beta)g_{\alpha_{n+1}}}{\sigma_{\ell}}\chi^{n+1;\sigma_{\ell}}_{\alpha_1\dots\alpha_n\alpha_{n+1}}(\bm{v})f(\bm{x})\;.
\end{equation}
We used for the $\chi$'s the conventions fixed in \S~\ref{ap:OU}, and we defined $\bm{\mathcal{M}}$ through (\ref{MA08p:diffBIS}).\\
The leading-order correction does not impose any solvability condition.
Since by (\ref{ap:OU_Green}) $\left(\mathcal{L}^{(0:0)}\right)^{-1}$ acts on $\chi^{n;\sigma_{\ell}}_{\dots}$
as a multiplication by $\tau_{\ell}/n$ (asymptotically in $t^{\ddag}$), we find
\begin{equation} \label{MA08p:1}
 p^{(0:1)}(\bm{x},\bm{v},t)=\tau_{\ell}\chi^{1;\sigma_{\ell}}_{\alpha_1}(\bm{v})\mathcal{M}_{\alpha_1}\xi^{(0:0)}(\bm{x},t)\;.
\end{equation}
The function $\xi^{(0:0)}$ in (\ref{MA08p:MB}) and (\ref{MA08p:1}) is specified by the solvability condition
brought about by (\ref{terza}), the solution whereof is
\begin{equation} \label{MA08p:2}
 p^{(0:2)}(\bm{x},\bm{v},t)=\chi^{0;\sigma_{\ell}}(\bm{v})\xi^{(0:2)}(\bm{x},t)+\frac{\tau_{\ell}^2}{2}\chi^{2;\sigma_{\ell}}_{\alpha_1\alpha_2}(\bm{v})\mathcal{M}_{\alpha_1}\mathcal{M}_{\alpha_2}\xi^{(0:0)}(\bm{x},t)
\end{equation}
if and only if $\xi^{(0:0)}$ satisfies the advection--diffusion equation
\begin{subequations}
 \begin{eqnarray} \label{MA08p:FP}
  \mathcal{N}\xi^{(0:0)}(\bm{x},t)=0\;,
 \end{eqnarray}
 \begin{eqnarray} \label{MA08p:FPinit}
  \xi^{(0:0)}(\bm{x},0)=\bar{\rho}(\bm{x})\;,
 \end{eqnarray}
\end{subequations}
with
\[\mathcal{N}:=\frac{\partial}{\partial t}+\bm{u}(\bm{x},t)\cdot\frac{\partial}{\partial\bm{x}}-\kappa\Delta_{\bm{x}}\equiv\frac{\partial}{\partial t}+\bm{u}(\bm{x},t)\cdot\frac{\partial}{\partial\bm{x}}-\frac{2\tau_{\ell}\sigma_{\ell}^2}{\mathrm{Pe}}\Delta_{\bm{x}}\;.\]
The solvability condition (\ref{MA08p:FP}) is the Fokker--Planck limit for \emph{incompressible}
advection of the Kramers dynamics $\mathcal{L}^{(0)}$ for vanishing Stokes.
As usual in the non-periodic case \citep{PS05}, the solvability condition arises
from the requirement that the non-homogeneous term in (\ref{terza})
be orthogonal to the kernel of the zero-frequency component of $\mathcal{L}^{(0:0)\dagger}$,
i.e.\ the generator of the Ornstein--Uhlenbeck process (\ref{ap:OU_gen}),
with respect to the scalar product on $\mathbb{L}^2(\mathbb{R}^{d})$ (equation (\ref{ap:OU_kernel})).
The fulfillment of this condition prevents the occurrence of marginal (polynomial in $t^{\ddag}$)
instabilities in the perturbative expansion.
By (\ref{MA08p:FP}), any integrable initial condition $\bar{\rho}_{x}$
on the periodicity cell $\mathbb{B}$ converges to a uniform distribution.
Since we are ultimately interested in degrees of freedom slower than the $t$-dynamics,
it is not restrictive to choose as initial condition the invariant measure
\begin{equation} \label{MA08p:uniform}
 \xi^{(0:0)}(\bm{x},t)=\bar{\rho}_{x}(\bm{x})=\frac{1}{\ell^d}\;.
\end{equation}
Finally, the presence in (\ref{MA08p:2}) of the homogeneous term $\chi^{0;\sigma_{\ell}}\xi^{(0:2)}$ takes
into account the solvability conditions intervening in the equation for $p^{(0:4)}$
(similar to (\ref{quarta}) with each superscript to $\bm{\psi}$ augmented by one). Namely,
parity considerations on the co-velocity distribution evince that
(\ref{quarta}) in its original form does not require any solvability condition and yields
\begin{eqnarray} \label{MA08p:3}
 \lefteqn{p^{(0:3)}=\frac{1-\beta}{\ell^d\mathrm{Fr}^2}\sqrt{\mathrm{Pe}}\,G_{\alpha_1}\chi^{1;\sigma_{\ell}}_{\alpha_1}+\tau_{\ell}\chi^{1;\sigma_{\ell}}_{\alpha_1}\mathcal{M}_{\alpha_1}\xi^{(0:2)}-\frac{\tau_{\ell}(1-\beta)}{\sigma_{\ell}\ell^d}\sqrt{\frac{\mathrm{Pe}}{2}}\,\chi^{1;\sigma_{\ell}}_{\alpha_1}\mathcal{L}_{(0:2)}u_{\alpha_1}}\nonumber\\
 &&+\frac{\tau_{\ell}^2(1-\beta)}{2\ell^d}\left[\sqrt{\frac{\mathrm{Pe}}{2}}\,\frac{\chi^{3;\sigma_{\ell}}_{\alpha_1\alpha_2\alpha_3}}{3\sigma_{\ell}}\mathcal{M}_{\alpha_3}-\left(\chi^{1;\sigma_{\ell}}_{\alpha_1}\frac{\partial}{\partial x_{\alpha_2}}+\chi^{1;\sigma_{\ell}}_{\alpha_2}\frac{\partial}{\partial x_{\alpha_1}}\right)\right]\mathcal{M}_{\alpha_2}u_{\alpha_1}\;,
\end{eqnarray}
having used (\ref{MA08p:uniform}) and
\[\mathcal{M}_{\alpha}\xi^{(0:0)}(\bm{x},t)=\frac{1-\beta}{\tau_{\ell}\sigma_{\ell}\ell^d}\sqrt{\frac{\mathrm{Pe}}{2}}\,u_{\alpha}(\bm{x},t)\;.\]
The complete expression of $p^{(0)}$ within third-order accuracy in $\mathrm{St}$
is then attained by applying (\ref{MA08p:L1recursion}) to (\ref{MA08p:3}), which
shows that $\xi^{(0:2)}$ satisfies the following advection--diffusion equation in $\mathbb{B}$ \citep{MA08}:
\begin{subequations}
 \begin{eqnarray} \label{MA08p:cc2}
  \mathcal{N}\xi^{(0:2)}(\bm{x},t)=\frac{\tau_{\ell}(1-\beta)}{\ell^d}\frac{\partial u_{\alpha_2}}{\partial x_{\alpha_1}}(\bm{x},t)\frac{\partial u_{\alpha_1}}{\partial x_{\alpha_2}}(\bm{x},t)\;,
 \end{eqnarray}
 \begin{eqnarray} \label{MA08p:cc3}
  \xi^{(0:2)}(\bm{x},0)=0\;.
 \end{eqnarray}
\end{subequations}
Owing to periodicity and incompressibility, the forcing sustaining (\ref{MA08p:cc2}) is zero average,
\[\int_{\mathbb{B}}\!\ud\bm{x}\,\frac{\partial u_{\alpha_2}}{\partial x_{\alpha_1}}(\bm{x},t)\frac{\partial u_{\alpha_1}}{\partial x_{\alpha_2}}(\bm{x},t)=-\int_{\mathbb{B}}\!\ud\bm{x}\,u_{\alpha_2}(\bm{x},t)\frac{\partial^2u_{\alpha_1}}{\partial x_{\alpha_1}\partial x_{\alpha_2}}(\bm{x},t)=0\;,\]
so ensuring probability conservation within the consistent accuracy:
\begin{equation} \label{add}
 \int_{\mathbb{B}}\ud\bm{x}\,\xi^{(0:2)}(\bm{x},t)=\!\mea\,p^{(0:2)}(\bm{x},\bm{v},t)=0\;.
\end{equation}

\subsection{Auxiliary vector field $\psi_{\mu}\equiv p^{(1)}_{\mu}$, $\mu=1,\dots,d$} \label{sec:psi}

The Stokes-number expansion of the auxiliary vector field $\bm{p}^{(1)}$ differs from that of
the phase-space density $p^{(0)}$ only by the presence of extra non-homogeneous terms.
These latter simplify when the coefficients (\ref{smin:vt_coe}) of the Stokes expansion of the
terminal velocity are evaluated using the results of the previous subsection for the phase-space
density. In particular we have
\[\bm{w}^{(:-1)}=\bm{w}^{(:1)}=0\;,\]
since the co-velocity dependence of the coefficients of the phase-space density expansion
is defined under parity
\begin{equation} \label{psi:parity}
 p^{(0:n)}(\bm{x},-\bm{v},t)=(-1)^np^{(0:n)}(\bm{x},\bm{v},t)
\end{equation}
and by (\ref{ap:OU_ortho}). By (\ref{labo}) we also have
\[\bm{w}^{(:0)}=\!\mea\,\left\{\bm{v}\frac{(1-\beta)\chi^{1;\sigma_{\ell}}_{\alpha_1}(\bm{v})u_{\alpha_1}(\bm{x},t)}{\sigma_{\ell}\ell^d}+\beta\bm{u}(\bm{x},t)\frac{\chi^{0;\sigma_{\ell}}(\bm{v})}{\ell^d}\right\}=0\;.\]
The first non-vanishing contribution is then
\[\bm{w}^{(:2)}=\!\mear\,\bm{u}(\bm{x},t)\xi^{(0:2)}(\bm{x},t)+\tau_{\ell}(1-\beta)\bm{g}\;,\]
using (\ref{ap:OU_ortho}) again and the vanishing of integrals (over the period) of derivatives of periodic functions.

\subsubsection{Leading orders of the Stokes expansion}
Proceeding as for the phase-space density, we look for a solution of (\ref{prima}) in the form
\begin{equation} \label{psi:0}
 p^{(1:0)}_{\mu}(\bm{x},\bm{v},t)=\chi^{0;\sigma_{\ell}}(\bm{v})\xi^{(1:0)}_{\mu}(\bm{x},t)\;.
\end{equation}
In full analogy to \S~\ref{sec:MA08p}, the vector field $\bm{\xi}^{(1:0)}$ is determined
by the first solvability condition brought about by the perturbative hierarchy.
The analogy with the phase-space density extends inasmuch that $\left(\mathcal{L}^{(0:0)}\right)^{-1}$ remains bounded
when acting on the non-homogeneous terms sustaining the first correction term in Stokes (\ref{seconda}), whence
\begin{equation} \label{psi:1}
 p^{(1:1)}_{\mu}(\bm{x},\bm{v},t)=\tau_{\ell}\chi^{1;\sigma_{\ell}}_{\alpha_1}(\bm{v})\mathcal{M}_{\alpha_1}\xi^{(1:0)}_{\mu}(\bm{x},t)-\frac{\tau_{\ell}\sigma_{\ell}}{\ell^d}\sqrt{\frac{2}{\mathrm{Pe}}}\,\chi^{1;\sigma_{\ell}}_{\mu}(\bm{v})\;.
\end{equation}
Thus, $\bm{\xi}^{(1:0)}$ is specified by the solvability condition associated to the second-order
equation (\ref{terza}). This latter differs from (\ref{MA08p:FP}) by the occurrence of
a non-homogeneous term on the right-hand side, and gives (\ref{csi0BIS}) complemented by the initial condition (\ref{zic}).\\
An important consequence of the very definition of terminal velocity
(\ref{vt}) preserved by its Stokes-number expansion (\ref{smin:vt}) is
the conservation law
\begin{equation} \label{zn}
 \int_{\mathbb{B}}\!\ud\bm{x}\,\bm{\xi}^{(1:0)}(\bm{x},t)=\!\mea\,\bm{p}^{(1:0)}(\bm{x},\bm{v},t)=\bm{0}\;.
\end{equation}
Inserting (\ref{csi0BIS}) into (\ref{terza}) yields
\begin{eqnarray} \label{psi:2}
 \lefteqn{p^{(1:2)}_{\mu}(\bm{x},\bm{v},t)=\chi^{0;\sigma_{\ell}}(\bm{v})\xi^{(1:2)}_{\mu}(\bm{x},t)}\nonumber\\
 &&+\frac{\tau_{\ell}}{2}\chi^{2;\sigma_{\ell}}_{\alpha_1\alpha_2}(\bm{v})\left[\tau_{\ell}\mathcal{M}_{\alpha_1}\mathcal{M}_{\alpha_2}\xi^{(1:0)}_{\mu}(\bm{x},t)-\frac{1-\beta}{\ell^d}\delta_{\alpha_1\mu}u_{\alpha_2}(\bm{x},t)\right]\;.
\end{eqnarray}
Like (\ref{MA08p:2}), (\ref{psi:2}) comprises a term $\chi^{0;\sigma_{\ell}}\,\bm{\xi}^{(1:2)}$
solution of the associated homogeneous equation. The introduction of such term is required
to impose the solvability condition in the equation for $\bm{p}^{(1:4)}$ (in which,
with respect to (\ref{quarta}), beyond the aforementioned augmentation of the superscripts,
an additional term $-w^{(:2)}_{\mu}\psi^{(:0)}_0$ appears within square brackets). Again, parity considerations
(\ref{psi:parity}) as in \S~\ref{sec:MA08p} entail that equation (\ref{quarta}) does not engender
any solvability condition. Tedious but straightforward calculations using (\ref{MA08p:L1recursion}),
(\ref{MA08p:L3recursion}) and (\ref{ap:OU_recursion}) yield
\begin{eqnarray*}
 \lefteqn{p^{(1:3)}_{\mu}=\frac{1-\beta}{\mathrm{Fr}^2}\sqrt{\mathrm{Pe}}\,G_{\alpha_1}\chi^{1;\sigma_{\ell}}_{\alpha_1}\xi^{(1:0)}_{\mu}+\tau_{\ell}\chi^{1;\sigma_{\ell}}_{\alpha_1}\mathcal{M}_{\alpha_1}\xi^{(1:2)}_{\mu}}\\
 &&-\tau_{\ell}^2\chi^{1;\sigma_{\ell}}_{\alpha_1}\mathcal{L}^{(0:2)}\mathcal{M}_{\alpha_1}\xi^{(1:0)}_{\mu}+\frac{\tau_{\ell}^3}{6}\chi^{3;\sigma_{\ell}}_{\alpha_1\alpha_2\alpha_3}\mathcal{M}_{\alpha_1}\mathcal{M}_{\alpha_2}\mathcal{M}_{\alpha_3}\xi^{(1:0)}_{\mu}\\
 &&-\frac{\tau_{\ell}^3\sigma_{\ell}}{2}\sqrt{\frac{2}{\mathrm{Pe}}}\,\left(\chi^{1;\sigma_{\ell}}_{\alpha_1}\frac{\partial}{\partial x_{\alpha_2}}+\chi^{1;\sigma_{\ell}}_{\alpha_2}\frac{\partial}{\partial x_{\alpha_1}}\right)\mathcal{M}_{\alpha_1}\mathcal{M}_{\alpha_2}\xi^{(1:0)}_{\mu}-\tau_{\ell}\sigma_{\ell}\sqrt{\frac{2}{\mathrm{Pe}}}\,\chi^{1;\sigma_{\ell}}_{\mu}\xi^{(0:2)}\\
 &&-\frac{\tau_{\ell}^2(1-\beta)}{2\ell^d}\chi^{1;\sigma_{\ell}}_{\alpha_1}\left(\mathcal{M}_{\alpha_1}u_{\mu}+\mathcal{M}_{\mu}u_{\alpha_1}\right)\\
 &&-\frac{\tau_{\ell}\beta(1-\beta)}{\sigma_{\ell}\ell^d}\sqrt{\frac{\mathrm{Pe}}{2}}\,\chi^{1;\sigma_{\ell}}_{\alpha_1}u_{\mu}u_{\alpha_1}+\frac{\tau_{\ell}^2\sigma_{\ell}}{2\ell^d}\sqrt{\frac{2}{\mathrm{Pe}}}\,\chi^{1;\sigma_{\ell}}_{\alpha_1}\frac{\partial}{\partial x_{\mu}}u_{\alpha_1}\;.
\end{eqnarray*}
The expression within third-order accuracy of $\bm{p}^{(1)}$ is complete observing that
$\bm{\xi}^{(1:2)}$ satisfies the solvability condition (\ref{psi:cc12BIS}) with initial condition (\ref{vic}),
and the conservation law
\[\int_{\mathbb{B}}\ud\bm{x}\,\bm{\xi}^{(1:2)}(\bm{x},t)=\!\mea\,\bm{p}^{(1:2)}(\bm{x},\bm{v},t)=\bm{0}\;.\]

\subsection{Expression of the eddy diffusivity} \label{sec:eda}

\subsubsection{Evaluation of the $O(\mathrm{St}^0)$ term}
Using (\ref{psi:0}), (\ref{psi:1}) and expressing the co-velocity
dependence in the integrands as the $\mathbb{L}^2(\mathbb{R}^{d})$-scalar
product between the Hermite polynomials $H^{n;\sigma_{\ell}}_{\dots}$ and their duals
$\chi^{n;\sigma_{\ell}}_{\dots}$, we have
\begin{eqnarray*}
 \lefteqn{K^{(:0)}_{\mu\nu}=-\mathcal{S}\!\mea\,\left[\sigma_{\ell}\sqrt{\frac{2}{\mathrm{Pe}}}\,H^{1;\sigma_{\ell}}_{\mu}(\bm{v})p^{(1:1)}_{\nu}+\beta u_{\mu}H^{0;\sigma_{\ell}}(\bm{v})p^{(1:0)}_{\nu}\right]}\\
 &=&-\mathcal{S}\!\mea\,H^{1;\sigma_{\ell}}_{\mu}(\bm{v})\left[\tau_{\ell}\sigma_{\ell}\sqrt{\frac{2}{\mathrm{Pe}}}\,\chi^{1;\sigma_{\ell}}_{\alpha_1}(\bm{v})\mathcal{M}_{\alpha_1}\xi^{(1:0)}_{\nu}-\frac{2\tau_{\ell}\sigma_{\ell}^2}{\ell^d\mathrm{Pe}}\chi^{1;\sigma_{\ell}}_{\nu}(\bm{v})\right]\\
 &&-\mathcal{S}\!\mea\,H^{0;\sigma_{\ell}}(\bm{v})\beta\chi^{0;\sigma_{\ell}}(\bm{v})u_{\mu}\xi^{(1:0)}_{\nu}\;.
\end{eqnarray*}
The orthogonality relations (\ref{ap:OU_ortho}), together with
\[\mear\,\frac{\partial\xi^{(1:0)}_{\nu}}{\partial x_{\alpha}}(\bm{x},t)=0\;,\]
yield (\ref{k0}). The positive definiteness of $K^{(:0)}_{\mu\nu}$ can be proven through (\ref{csi0BIS}) and
\[\mear\,\left[\frac{\partial}{\partial t}+\bm{u}(\bm{x},t)\cdot\frac{\partial}{\partial\bm{x}}\right]\xi^{(1:0)}_{\mu}(\bm{x},t)\xi^{(1:0)}_{\nu}(\bm{x},t)=0\;.\]

\subsubsection{Evaluation of the $O(\mathrm{St}^1)$ term}
The leading-order correction
\[K^{(:2)}_{\mu\nu}=-\mathcal{S}\!\mea\,\left[\sigma_{\ell}\sqrt{\frac{2}{\mathrm{Pe}}}\,H^{1;\sigma_{\ell}}_{\mu}(\bm{v})p^{(1:3)}_{\nu}+\beta u_{\mu}H^{0;\sigma_{\ell}}(\bm{v})p^{(1:2)}_{\nu}\right]\]
can be evaluated as the sum of two distinct integrals.\\
The first involves the projection of $p^{(1:3)}_{\nu}$ over $H^{1;\sigma_{\ell}}_{\mu}$. Using (\ref{ap:OU_ortho})
to integrate out the co-velocity dependence, we are left with
\begin{eqnarray} \label{kappa:21}
 \lefteqn{-\mathcal{S}\!\mea\,\sigma_{\ell}\sqrt{\frac{2}{\mathrm{Pe}}}\,H^{1;\sigma_{\ell}}_{\mu}(\bm{v})p^{(1:3)}_{\nu}}\nonumber\\
 &=&-\mathcal{S}\!\mear\,\left[\sqrt{2}\,\frac{\sigma_{\ell}(1-\beta)}{\mathrm{Fr}^2}G_{\mu}\xi^{(1:0)}_{\nu}+\tau_{\ell}\sigma_{\ell}\sqrt{\frac{2}{\mathrm{Pe}}}\,\left(\mathcal{M}_{\mu}\xi^{(1:2)}_{\nu}-\tau_{\ell}\mathcal{L}^{(0:2)}\mathcal{M}_{\mu}\xi^{(1:0)}_{\nu}\right)\right]\nonumber\\
 &&+\mathcal{S}\!\mear\,\left[\frac{\tau_{\ell}^3\sigma_{\ell}^2}{\mathrm{Pe}}\,\frac{\partial}{\partial x_{\alpha}}\left(\mathcal{M}_{\mu}\mathcal{M}_{\alpha}+\mathcal{M}_{\alpha}\mathcal{M}_{\mu}\right)\xi^{(1:0)}_{\nu}+\sqrt{\frac{2}{\mathrm{Pe}}}\,\frac{\tau_{\ell}^2\sigma_{\ell}(1-\beta)}{2\ell^d}\mathcal{M}_{\nu}u_{\mu}\right]\nonumber\\
 &&+\!\mear\,\left[\frac{2\tau_{\ell}\sigma_{\ell}^2}{\mathrm{Pe}}\delta_{\mu\nu}\xi^{(0:2)}+\frac{\tau_{\ell}\beta(1-\beta)}{\ell^d}u_{\mu}u_{\nu}-\frac{\tau_{\ell}^2\sigma_{\ell}^2}{\ell^d\mathrm{Pe}}\,\mathcal{S}\frac{\partial}{\partial x_{\nu}}u_{\mu}\right]\nonumber\\
 &=&-\mathcal{S}\!\mear\,\left[(1-\beta)u_{\mu}\xi^{(1:2)}_{\nu}-\frac{\tau_{\ell}(1-\beta)}{\ell^d}u_{\mu}u_{\nu}\right]\;,
\end{eqnarray}
where the latter simplifications are due to the conservation laws (\ref{add}) and (\ref{zn}), combined with the vanishing of
integrals (over the period) of derivatives of periodic functions.\\
The second term involves (\ref{psi:2}) and is more straightforward to evaluate:
\begin{equation} \label{kappa:22}
 -\mathcal{S}\!\mea\,\beta u_{\mu}H^{0;\sigma_{\ell}}(\bm{v})p^{(1:2)}_{\nu}=-\!\mear\,\frac{\beta}{2}\left(u_{\mu}\xi^{(1:2)}_{\nu}+u_{\nu}\xi^{(1:2)}_{\mu}\right)\;.
\end{equation}
Expressions (\ref{kappa:21}) and (\ref{kappa:22}) combine into (\ref{kappa:2}),
which can be gleaned with (\ref{kappa:0}) to give the final result (\ref{kappa:final}).

\subsection{Expansion at small inertia and finite terminal velocity} \label{sec:app}

The leading-order correction in Stokes to the phase-space density becomes
\begin{equation} \label{Ma:p1}
 p^{(0:1)}\mapsto\tilde{p}^{(0:1)}(\bm{x},\bm{v},t)=\tau_{\ell}\chi^{1;\sigma_{\ell}}_{\alpha_1}(\bm{v})\tilde{\mathcal{M}}_{\alpha_1}\tilde{\xi}^{(0:0)}(\bm{x},t)\;,
\end{equation}
where now $\tilde{\xi}^{(0:0)}$ satisfies
\begin{equation} \label{Maxey:sc00}
 \tilde{\mathcal{N}}\tilde{\xi}^{(0:0)}(\bm{x},t)=0\;,
\end{equation}
with
\[\mathcal{N}\mapsto\tilde{\mathcal{N}}:=\frac{\partial}{\partial t}+\left[\bm{u}+(1-\beta)\bm{g}\tau_{\mathrm{S}}\right]\cdot\frac{\partial}{\partial\bm{x}}-\kappa\Delta_{\bm{x}}\equiv\frac{\partial}{\partial t}+\left[\bm{u}+(1-\beta)S_FU\bm{G}\right]\cdot\frac{\partial}{\partial\bm{x}}-\frac{U\ell}{\mathrm{Pe}}\Delta_{\bm{x}}\;.\]
The new solvability condition (\ref{Maxey:sc00}) remains however compatible with the
steady-state solution (\ref{MA08p:uniform}), which we therefore assume. A further consequence
of (\ref{Ma:p1}) is that the first non-vanishing contribution to the terminal velocity
occurs at $O(\mathrm{St}^0)$:
\[\bm{w}^{(:0)}\mapsto\tilde{\bm{w}}^{(:0)}=(1-\beta)S_FU\bm{G}\;.\]
The leading-order correction to $\bm{p}^{(1)}$ on its turn becomes
\[\bm{p}^{(1:1)}\mapsto\tilde{\bm{p}}^{(1:1)}(\bm{x},\bm{v},t)=\tau_{\ell}\chi^{1;\sigma_{\ell}}_{\alpha_1}(\bm{v})\tilde{\mathcal{M}}_{\alpha_1}\tilde{\bm{\xi}}^{(1:0)}(\bm{x},t)-\frac{\tau_{\ell}\sigma_{\ell}}{\ell^d}\sqrt{\frac{2}{\mathrm{Pe}}}\,\bm{\chi}^{1;\sigma_{\ell}}\;,\]
where now $\tilde{\bm{\xi}}^{(1:0)}$ obeys (\ref{csitilde}) supplemented with (\ref{cin}).
Correspondingly, throughout this section (\ref{MA08p:diffBIS}) becomes
\[\bm{\mathcal{M}}\mapsto\tilde{\bm{\mathcal{M}}}:=\frac{(1-\beta)}{\tau_{\ell}\sigma_{\ell}}\sqrt{\frac{\mathrm{Pe}}{2}}\,\left[\bm{u}(\bm{x},t)+S_FU\bm{G}\right]-\sqrt{\frac{2}{\mathrm{Pe}}}\,\sigma_{\ell}\frac{\partial}{\partial\bm{x}}\;.\]

\section{Ornstein--Uhlenbeck process} \label{ap:OU}

We briefly recall some properties of the well-known Ornstein--Uhlenbeck
process \citep[see, e.g.,][and references therein]{V07} used in the main text.
The generator of the Ornstein--Uhlenbeck process on $\mathbb{R}^{d}$ with
parametres $\tau,\sigma>0$, defined as
\begin{equation} \label{ap:OU_gen}
 \mathcal{G}_{\tau,\sigma}^{\dagger}:=-\frac{\bm{y}}{\tau}\cdot\frac{\partial}{\partial\bm{y}}+\frac{\sigma^2}{\tau}\Delta_{\bm{y}}
\end{equation}
is diagonal on the basis of the multidimensional Hermite polynomials $H^{n;\sigma}_{\alpha_1\dots\alpha_n}$:
\[\mathcal{G}_{\tau,\sigma}^{\dagger}H^{n;\sigma}_{\alpha_1\dots\alpha_n}=-\frac{n}{\tau}H^{n;\sigma}_{\alpha_1\dots\alpha_n}\;,\]
with the $1\leq\alpha_i\leq d$ ($i=1,\dots,n$) indices ranging over the $\bm{y}$-vector components.
Hermite (multi-)polynomials are most conveniently defined by the Rodrigues formula
\[H^{n;\sigma}_{\alpha_1\dots\alpha_n}(\bm{y})=(-1)^n\frac{\sigma^n}{\chi^{0;\sigma}(\bm{y})}\left(\prod_{i=1}^n\frac{\partial}{\partial y_{\alpha_i}}\right)\chi^{0;\sigma}(\bm{y})\;,\]
where
\[\chi^{0;\sigma}(\bm{y})=\frac{\ue^{-y^2/2\sigma^2}}{(2\pi\sigma^2)^{d/2}}\]
($y^2\equiv||\bm{y}||^2$) is the unique element
of the kernel (steady-state measure) of the adjoint $\mathcal{G}_{\tau,\sigma}$
operator on $\mathbb{L}^2(\mathbb{R}^{d})$:
\begin{equation} \label{ap:OU_dual}
 \mathcal{G}_{\tau,\sigma}=\frac{\partial}{\partial\bm{y}}\cdot\frac{\bm{y}}{\tau}+\frac{\sigma^2}{\tau}\Delta_{\bm{y}}=\chi^{0;\sigma}\mathcal{G}_{\tau,\sigma}^{\dagger}\frac{1}{\chi^{0;\sigma}}\;.
\end{equation}
The last equality in (\ref{ap:OU_dual}) implies that $\mathcal{G}_{\tau,\sigma}$ is diagonal,
\[\mathcal{G}_{\tau,\sigma}\chi^{n;\sigma}_{\alpha_1\dots\alpha_n}=-\frac{n}{\tau}\chi^{n;\sigma}_{\alpha_1\dots\alpha_n}\;,\]
on the $\mathbb{L}^2(\mathbb{R}^{d})$ complete set
\begin{equation} \label{ap:OU_basis}
 \chi^{n;\sigma}_{\alpha_1\dots\alpha_n}=H^{n;\sigma}_{\alpha_1\dots\alpha_n}\chi^{0;\sigma}\;.
\end{equation}
We also use the convention
\[\chi^{n;\sigma}_{\alpha_1\dots\alpha_n}=H^{n;\sigma}_{\alpha_1\dots\alpha_n}=0\,,\hspace{1.0cm}\textrm{if}\hspace{0.2cm}n<0\;.\]
The Rodrigues formula implies the $\mathbb{L}^{2}(\mathbb{R}^{d})$-orthogonality
relations between Hermite polynomials and the basis elements (\ref{ap:OU_basis}).
Namely, a multiple integration by parts
readily shows that the integral is non-vanishing only for $n_1=n_2=n$,
and if there exists at least one permutation $\mathcal{P}$ of the $n$-tuple $(\beta_1,\dots,\beta_n)$
such that $\alpha_i=\mathcal{P}_i(\beta_1,\dots,\beta_n)$:
\begin{eqnarray} \label{ap:OU_ortho}
 \int_{\mathbb{R}^d}\!\ud\bm{y}\,H^{n_1;\sigma}_{\alpha_1\dots\alpha_{n_1}}\chi^{n_2;\sigma}_{\beta_1\dots\beta_{n_2}}&=&\sigma^{n_1+n_2}\!\int_{\mathbb{R}^{d}}\!\ud\bm{y}\,\chi^{0;\sigma}\left(\prod_{i_2=1}^{n_2}\frac{\partial}{\partial y_{\beta_{i_2}}}\right)\left[\frac{(-1)^{n_1}}{\chi^{0;\sigma}}\left(\prod_{i_1=1}^{n_1}\frac{\partial}{\partial y_{\alpha_{i_1}}}\right)\chi^{0;\sigma}\right]\nonumber\\
 &=&\delta_{n_1n_2}\sum_{\{\mathcal{P}\}}\prod_{i=1}^n\delta_{\alpha_i,\mathcal{P}_i(\beta_1,\dots,\beta_n)}
\end{eqnarray}
(the sum ranging over all permutations of the $(\beta_1\dots\beta_{n})$).
Further consequences are the grading relation
\begin{equation} \label{ap:OU_grading}
 \frac{\partial\chi^{n;\sigma}_{\alpha_1\dots\alpha_n}}{\partial y_{\alpha_{n+1}}}=-\frac{\chi^{n+1;\sigma}_{\alpha_1\dots\alpha_n\alpha_{n+1}}}{\sigma}
\end{equation}
and the recursion relation
\begin{equation} \label{ap:OU_recursion}
 H^{n+1;\sigma}_{\alpha_1\dots\alpha_n\alpha_{n+1}}(\bm{y})=\frac{y_{\alpha_{n+1}}}{\sigma}H^{n;\sigma}_{\alpha_1\dots\alpha_n}(\bm{y})-\sum_{k=1}^nH^{n-1;\sigma}_{\alpha_1\dots\alpha_{k-1}\alpha_{k+1}\dots\alpha_n}(\bm{y})\delta_{\alpha_k\alpha_{n+1}}\;,
\end{equation}
which follows by applying (\ref{ap:OU_ortho}) and (\ref{ap:OU_grading}) to
\[\frac{\partial H^{n;\sigma}_{\alpha_1\dots\alpha_n}}{\partial x_{\alpha_{n+1}}}(\bm{y})=-\frac{H^{n+1;\sigma}_{\alpha_1\dots\alpha_n\alpha_{n+1}}(\bm{y})}{\sigma}+\frac{y_{\alpha_{n+1}}}{\sigma^2}H^{n;\sigma}_{\alpha_1\dots\alpha_n}(\bm{y})\;.\]
Finally we observe that
\begin{eqnarray} \label{ap:OU_Green}
 \lefteqn{\left[\left(\mathcal{L}^{(0:0)}\right)^{-1}\chi^{n;\sigma_{\ell}}_{\alpha_1\dots\alpha_n}\right](\bm{y},t^{\star};t_o)}\nonumber\\
 &&=\int_{t_o}^{t^{\star}}\!\ud s\int_{\mathbb{R}^d}\ud\bm{z}\,\ue^{(t^{\star}-s)\mathcal{G}_{\tau_{\ell},\sigma_{\ell}}}(\bm{y},\bm{z})\chi^{n;\sigma_{\ell}}_{\alpha_1\dots\alpha_n}(\bm{z})=\chi^{n;\sigma_{\ell}}_{\alpha_1\dots\alpha_n}(\bm{y})\!\int_{t_o}^{t^{\star}}\!\ud s\,\ue^{-(t^{\star}-s)n/\tau_{\ell}}\nonumber\\
 &&\xrightarrow{t^{\star}\to\infty}\frac{\tau_{\ell}}{n}\chi^{n;\sigma_{\ell}}_{\alpha_1\dots\alpha_n}(\bm{y})\;,
\end{eqnarray}
which entails that $\left[\left(\mathcal{L}^{(0:0)}\right)^{-1}f\right](\bm{y},t;t_o)$ for any integrable
$f(\bm{y})$ is bounded in the time $t$ dependence if and only if
\begin{equation} \label{ap:OU_kernel}
 \int_{\mathbb{R}^d}\!\ud\bm{y}\,H^{0;\sigma}(\bm{y})f(\bm{y})\equiv\!\int_{\mathbb{R}^{d}}\!\ud\bm{y}\,f(\bm{y})=0\;.
\end{equation}

\end{document}